\documentclass[12pt,preprint]{aastex}

\newcommand{\dij}{}

\newcommand{\Ha}{\mbox{H$\alpha$}}
\newcommand{\sii}{S~{\sc ii}}

\def\c18o{C$^{18}$O($1\rightarrow 0$)}

\begin{document}


\title{Large Area Mapping at 850 Microns. 
V.  Analysis of the Clump Distribution in the Orion A South Molecular Cloud}

\author{Doug Johnstone\altaffilmark{1,2} \& John Bally\altaffilmark{3}}

\altaffiltext{1}
{National Research Council Canada, 
Herzberg Institute of Astrophysics,
5071 West Saanich Rd, Victoria, BC, V9E 2E7, Canada; 
doug.johnstone@nrc-cnrc.gc.ca}
\altaffiltext{2}{Department of Physics \& Astronomy, University of Victoria,
Victoria, BC, V8P 1A1, Canada}
\altaffiltext{3}
{Center for Astrophysics and Space Astronomy and Department of Astrophysical
and Planetary Sciences, University of Colorado, Boulder, CO 80309; 
bally@casa.colorado.edu}

\begin{abstract}

\noindent
We present results from a 2300 arcmin$^2$ survey of the Orion A
molecular cloud at 450 and 850\,$\mu$m 
using the Submillimetre Common-User Bolometer
Array (SCUBA) on the James Clerk Maxwell Telescope.  The region mapped lies
directly south of the OMC1 cloud core and includes OMC4, OMC5, HH1/2, HH34, 
and L1641N.  We identify 71 independent clumps in the 850\,$\mu$m map
and compute size, flux, and degree of central concentration in each. Comparison
with isothermal, pressure-confined, self-gravitating Bonnor-Ebert spheres
implies that the clumps have internal temperatures $T_d \sim 22 \pm 5\,$K
and surface pressures ${\rm log}(k^{-1}\,P\,{\rm cm}^{-3}\,{\rm K}) 
= 6.0 \pm 0.2$.  The clump masses span the range $0.3 - 22\,M_\odot$ assuming 
a dust temperature $T_d \sim 20\,$K and a dust
emissivity $\kappa_{850} = 0.02\,$cm$^{2}\,$g$^{-1}$.  The distribution of
clump masses is well characterized by a power-law $N(M) \propto M^{-\alpha}$
with $\alpha = 2.0 \pm 0.5$ for $M > 3.0\,M_\odot$, indicating a clump mass function
steeper than the stellar Initial Mass Function. Significant incompleteness
makes determination of the slope at lower masses difficult. 
A comparison of the submillimeter emission map with an H$_2$ 2.122\,$\mu$m
survey of the same region is performed. Several new Class 0 sources are
revealed and a correlation is found between both the column density and
degree of concentration of the submillimeter sources and the likelihood
of coincident H$_2$ shock emission.
\end{abstract}


\section{Introduction}

Star formation takes place within molecular clouds, in regions where 
the local pull of gravity overwhelms the random and ordered motions 
of the gas. Thus, in order to understand the physical process of star
formation it is necessary to observe the formation and evolution of
structure inside these clouds. The advent of sensitive millimeter
and submillimeter continuum array detectors has provided an opportunity
for large area mapping of the optically thin emission from cold dust
within molecular clouds, yielding observations of the column density 
distribution within these regions. The most common features observed in these
maps are dense condensations of dust, ``clumps"\footnote{Unfortunately,
the words ``cores" and ``clumps" have ambiguous meaning and are used
to denote various types of objects by different authors. In this paper
``clumps" refer to stellar-mass sized condensations of dust and gas.}.
These clumps reside in areas where stars are either forming or are expected
to form. When combined with observations being taken at shorter wavelengths  
by the Spitzer Space Telescope, it should be possible to unambiguously 
classify these objects as either 
starless (possibly prestellar) or protostellar (Class 0 or Class I)
(Jorgensen et al. 2006). 
Alternatively, comparison with other evolutionary tracers, 
such as H$_2$ shock emission (Walawender et al.  2005, 2006), provides 
evidence for which clumps already contain forming stars.

Continuing an effort to quantify the necessary pre-conditions for star 
formation, and in anticipation of the Spitzer observations,
in this paper we present submillimeter dust continuum maps of the 
Orion A South region, including OMC4, OMC5, HH1/2, HH34, and L1641N. 
Previous papers in the series investigated the northern portion
of the Orion A cloud containing the Orion Nebula (Johnstone \& Bally 
1999), the  Ophiuchus cloud (Johnstone et al. 2000b, hereafter Paper II), 
the Orion B North region (Johnstone et al. 2001, hereafter Paper III), 
and the Orion B South region (Johnstone, Matthews, \& Mitchell 2006, 
hereafter Paper IV).

The Orion A molecular cloud south of the Orion Nebula contains 
several active sites of ongoing star and star cluster formation. 
The region dubbed OMC4 is located about 10\arcmin\ south of the
Orion Nebula and consists of a V-shaped cluster of 
cores located near the top portion of the SCUBA survey. 
The double star $\iota$ Orionis (spectral types O9III, B1III),
about 30\arcmin\ south of the Orion Nebula, is the brightest member 
of the $>$3 Myr old NGC 1980 cluster located in the 
southern part of the Integral Shaped Filament (ISF) in Orion A.  
The $\iota$ Orionis system has been implicated in a four-body 
dynamical interaction that launched the famous high-velocity 
run-away stars AE Auriga and $\mu$ Columbae 2.6 Myr ago 
(Hoogerwerf, de Bruijne,  \& de Zeeuw 2001).  This portion of the 
Orion OB1c association must lie at least 10 pc in front of the 
Orion A cloud and must represent a considerably older site of star 
birth than either the Orion Nebula or the other currently active star 
forming regions in Orion A (Brown et al. 1994).

The embedded L1641N cluster contains at least 30 mid-infrared sources 
detected by ISO (Ali \& Noriega-Crespo 2004; Chen \& Tokunaga 1994).  
Young stars in this cluster power dozens of Herbig-Haro (HH) flows, including 
two giant flows that can be traced for several parsecs from the L1641 
cluster core (Reipurth, Devine, \& Bally 1998).  The L1641N cluster is
the most active site of on-going star formation in the fields covered
by this SCUBA survey.   

The western side of the Orion A cloud near the L1641N cluster contains
a number of Herbig-Haro objects and young stars not associated
with rich clusters.    The HH 34 jet, located about 16\arcmin\ west of 
the L1641 region, emerges from a star embedded within a small cloud 
which hosts a small group of young stars (e.g. Devine et. al.
1997).    The source and true nature of the bright but enigmatic 
bow-shaped HH~222,  located about 5\arcmin\ north-northwest of HH 34, 
remains unknown (Castets, Reipurth, \& Loinard, 2004).  However, another 
giant bow, HH~401, located about 5\arcmin\ southwest of HH 34 appears to 
trace a parsec-scale outflow possibly powered by the HH~1/2 jet (Ogura 1995). 

Near the southern periphery, our survey field contains the bright 
NGC1999 reflection nebula and its exciting star V380 Ori 
(spectral type A0e).  Many molecular outflows and Herbig-Haro objects
are observed here, suggesting that this region also
contains a small embedded group of young stars (Chen \& Tokunaga 1994).
NGC1999 is associated with a 
dense cloud core and molecular ridge, seen in millimeter CO and CS transitions, 
which extends about 10\arcmin\ towards the south (Morgan et al. 1991).
The southern end of this ridge contains the famous Herbig-Haro
objects HH~1, 2, and 3, as well as a number of fainter shocks (e.g.
HH 144/145 and HH~147).  This region contains another compact group of 
young stars.

\section{Observations and Data Reduction}\label{s_odr}

The data covering the Orion A South region were obtained between 1997 and
2002 using SCUBA at the JCMT. The JCMT archive hosted by the 
CADC\footnote{Extensive 
use of the JCMT archive at the Canadian Astronomy Data
Centre (CADC) was used to retrieve all relevant SCUBA data in Orion A South. 
The CADC is operated by the Dominion Astrophysical Observatory for the 
National Research Council of Canada's Herzberg Institute of Astrophysics.}
was used to find all relevant observations toward the region.
A total of 305 individual scan-maps, each approximately 100 arcmin$^2$,
were obtained, covering a combined 2300 square arcminute region.  
Each individual 3-arcsecond cell in the map was measured approximately 40
times at 850\,$\mu$m and 90 times at 450\,$\mu$m. Thus, the total time spent
observing each cell was approximately 5 and 11 seconds respectively.
{The weather conditions varied significantly during the many nights of 
observing, especially at 450$\,\mu$m. The mean and deviation of the
 optical depths at 
850$\,\mu$m were $\tau_{850} = 0.26 \pm 0.10$ and at 450$\,\mu$m
were $\tau_{450} = 1.1 \pm 0.3$.}

The raw bolometer data were reduced using the standard SCUBA software 
(Holland et al. 1999) to flat-field, extinction-correct, and
flux calibrate.  
{The extinction corrections, which provide atmospheric optical
depth corrections}, and flux
calibrations as a function of time were obtained from the JCMT calibrations 
archive (Jenness et al. 2002).  The extinction corrections are tabulated as 
a function of time based on a comparison
of skydip measurements taken with the JCMT during normal observing and zenith 
optical depth measures taken every few minutes at 225 GHz by the CSO tau meter.
The flux calibration values are tabulated from all calibration measurements
taken during extended periods of time, e.g. semesters, over which the 
telescope and instrument were not known to have changed significantly
(Jenness et al. 2002).
Typical uncertainties in the flux calibration can be determined from 
Jenness et al. (2002; their Figures 1 and 2). 
At 850\,$\mu$m the uncertainty 
is approximately $\pm10\,$\% while at 450$\,\mu$m the uncertainty is closer 
to $\pm25\,$\%. These values are consistent with  the uncertainties determined
by individual observers using calibrator sources.  Most of the uncertainty is 
caused by changes in the telescope surface due to temperature and gravity 
during the night and resultant changes in the beam profile, which can be quite 
significant at 450\,$\mu$m. 
{In addition, the beam profiles are not entirely Gaussian, resulting
in complications when attempting to determine total flux measurements for
extended objects. At 450$\,\mu$m this is a rather severe problem as a
significant portion of the total flux is contained in the broader error
beam.}

\begin{figure}[htp]
\includegraphics[width=0.49\textwidth,clip]{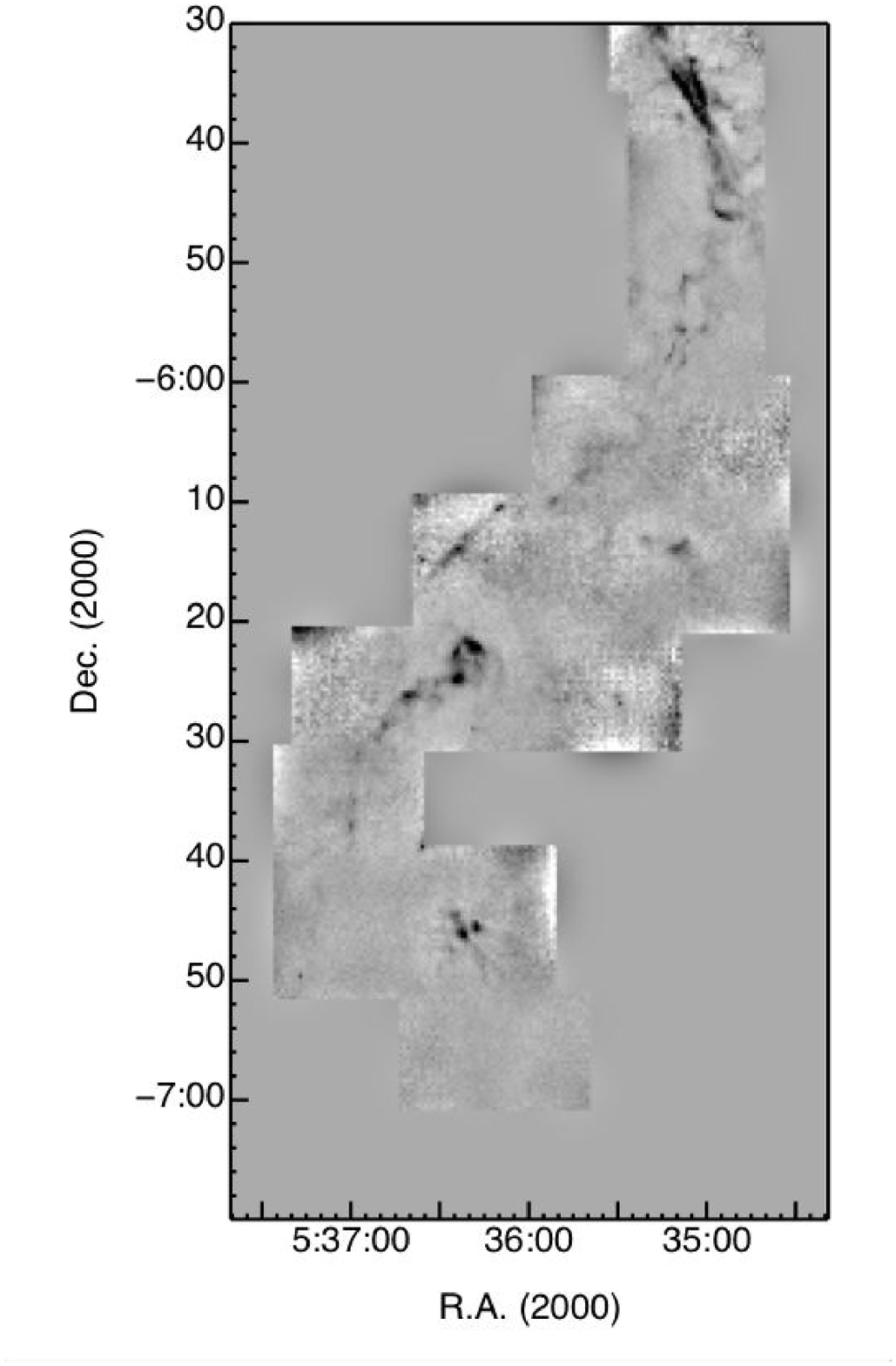}
\hfill
\includegraphics[width=0.49\textwidth,clip]{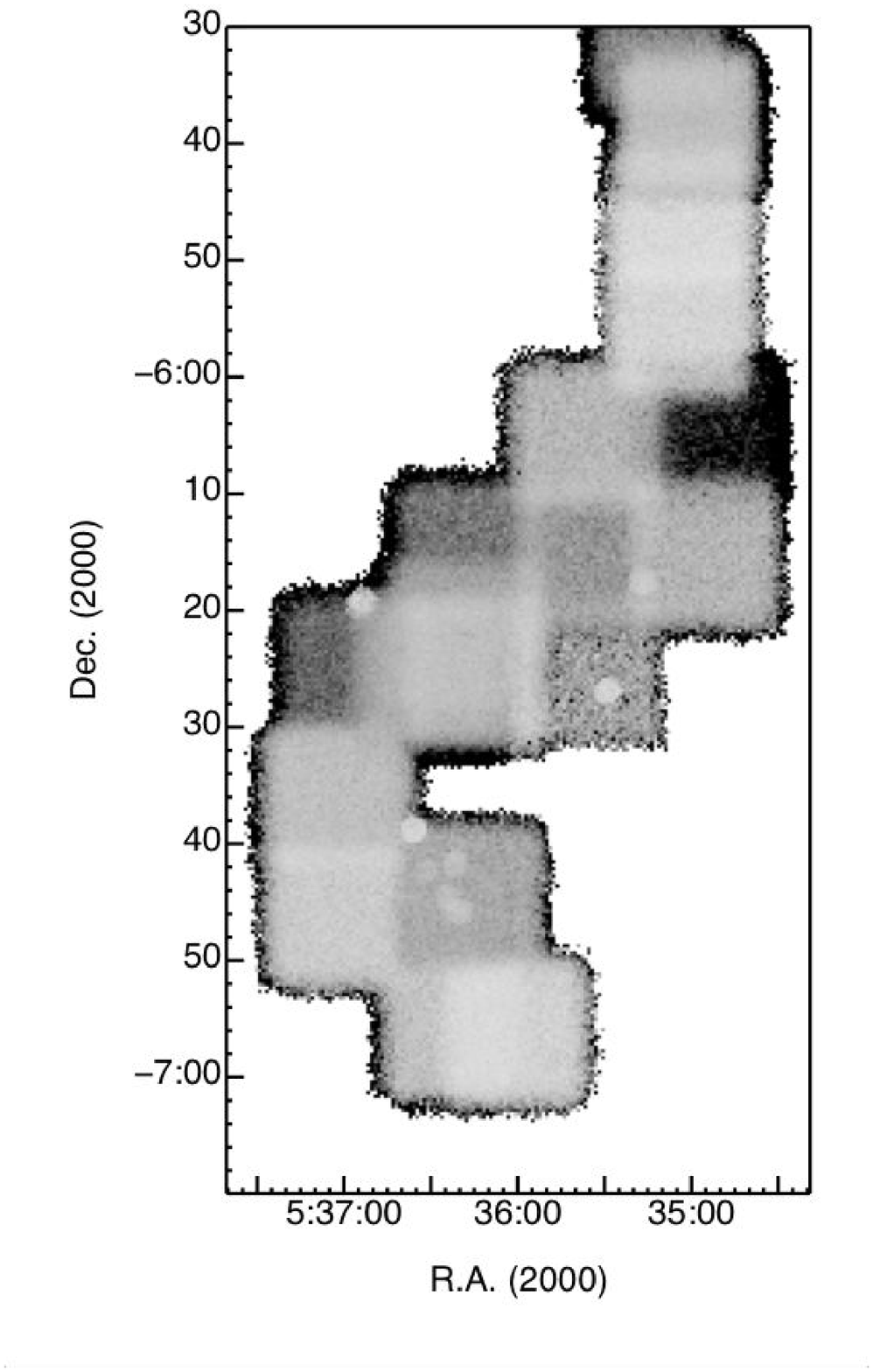}\\
\caption{(a) The Orion A South region at  850$\,\mu$m. The star forming
regions discussed in the text are labeled.
Left: 850$\,\mu$m emission (greyscale: white to black represents
-0.25 to 0.5 Jy\,bm$^{-1}$). Right: 850$\,\mu$m noise map (greyscale: white to
black represents 0 to 0.15 Jy\,bm$^{-1}$).  Note that the noise is
systematically lower in regions where multiple observations were taken.
(b)The Orion A South region at 450$\,\mu$m.
Left: 450$\,\mu$m emission (greyscale: white to black represents
-1 to 2 Jy\,bm$^{-1}$). Right: 450$\,\mu$m noise map
(greyscale: white to black represents 0 to 1  Jy\,bm$^{-1}$).
}\label{f_oas_850_450}
\end{figure}

The data were further reduced and transformed into
the final maps using the matrix inversion technique discussed by
Johnstone et al. (2000a, hereafter Paper I). 
This inversion technique has many advantages, in particular the ability
to utilize {\it all} data taken with SCUBA at the JCMT, regardless of the
observing method (i.e. arbitrary chop throws and chop directions)
and to properly weight each observation {by the noise inherent in the 
individual bolometer}.  Significant low-amplitude, 
large-scale features remain after the image reconstruction. 
Although these features may be real, artificial structure on these 
scales are often produced by weather variations and bolometer drift
since reconstruction of chopped data amplifies long spatial wavelength noise.
To filter these features, the map was convolved with a large Gaussian
($\sigma = 135''$) and the resultant smoothed map was subtracted from the
reconstruction.  In order to minimize the effect of over-subtraction in regions
with bright sources, all pixels greater than 0.25\,Jy\,bm$^{-1}$ 
{at 850\,$\mu$m and 5\,Jy\,bm$^{-1}$ at 450\,$\mu$m}
 were removed in the construction of the smoothed map.
The final, reconstructed 850$\,\mu$m and 450$\,\mu$m maps,
after filtering, are shown in Figure \ref{f_oas_850_450}
along with maps of the estimated uncertainty at each position. 
{Figure 2 shows blow ups of the four most interesting regions.} 
Note that the uncertainty in the measurement drops significantly in 
regions where multiple observations were obtained. 
{Although the noise in the final reconstructed maps is not uniform
across the image, the typical value per $3\arcsec$ pixel is 
0.03\,Jy\,bm$^{-1}$ at 850$\,\mu$m and 0.3\,Jy\,bm$^{-1}$ at 450$\,\mu$m.}

\section{Data Analysis: Clump Properties in Orion A South}\label{s_clumps}

\subsection{Identification of Clumps}

Clumps within the Orion A South region were identified in the
same manner used for Ophiuchus (Paper II), Orion B North (Paper III), 
and Orion B South (Paper IV), allowing for
a direct comparison of the results. Briefly, the clump finding algorithm,
{\tt clfind} (Williams et al. 1994), was employed to determine the location
and extent of individual clumps. This technique separates the map into
individual clumps by searching for local minima boundaries. Each 
pixel, above a limiting threshold, is assigned to a particular clump. 
Thus, unlike Gaussian clump-finding algorithms, the regularity of the clump
is not preset and clumps are allowed to have arbitrary shape. We 
remind the reader that any clump
finding algorithm introduces biases into the results through its assumption
of what constitutes an individual source; however, by applying the
same technique to an ensemble of regions, observed under similar conditions,
we intend to use the results of the analysis to highlight the similarities
and differences between star-forming regions. Thus, the exact nature of the
clump finding algorithm, although important for understanding the details of
the objects which it identifies, is not critical. We prefer {\tt clfind}
over other techniques, such as Gaussian-profile clump finding algorithms
(Stutzki \& G\"usten 1990), explicitly because it does not prejudge the
degree of isotropy of each region.

Initially 83 clumps were obtained by the {\tt clfind} routine; however,
12 were removed on visual inspection of the clump location and the
Orion A South map.  All clumps removed were near the edge of the map where
the residual noise is highest.  Following the earlier papers in this series, 
the effective radius
of each clump was determined by taking the total area enclosed within the
clump boundary and computing the radius of a circle required to reproduce
the result. The properties of the 71 identified clumps are presented in 
Table \ref{t_850data}. The total flux for each clump has not been corrected 
for the effect of the telescope error beam, in keeping with previous papers 
in this series.  The error beam at 850\,$\mu$m is at most $\sim 10$\%, 
{similar to the calibration uncertainties and thus the uncertainty
in the final derived total flux is less than 20\%.}
Figure \ref{f_oas_clumps} overlays circles on the Orion A South
map with the central location marking the measured center of each clump and
the circle size a measure of the extent of the clump.  It is clear from
visual inspection that in a few crowded regions the measured clumps are
asymmetric and may be composite objects.  
 
The mass for each clump was estimated by calculating the total flux from
within the clump boundary and assuming that the measured flux was due to
thermal emission from optically thin dust particles. As in Paper
IV, the opacity per unit 
mass column density at $\lambda = 850\,\mu$m was taken to be 
$\kappa_{850} = 0.02\,$cm$^{2}\,$g$^{-1}$. This value is larger than
the value used in Papers II and III,
{thus lowering the mass determination for a fixed total flux,}
 but reflects a consensus view that 
the dust in dense regions has a higher emissivity at long wavelengths 
(van der Tak et al. 1999).  We remind the reader that the
value of $\kappa$ depends sensitively on the properties of the dust
grains (see Henning, Michel, \& Stognienko 1995 for a review). 
{It is also possible for the continuum flux to be contaminated
by CO 3--2 emission; however, as shown by Johnstone et al. (2003)
this is not typically a concern within nearby star-forming regions
since the CO emission line needs to be both strong {\it and} 
extremely broad.}

Taking the distance to Orion A South to be $d = 450\,$pc 
the flux enclosed within the
clump boundary $S_{850}$, measured in Janskys, is converted to mass via
\begin{equation}
M_{\rm clump} = 0.75 \times S_{850}
\left[ \exp\left({17\,{\rm K} \over T_d}\right) -1\right]\,
\left( {\kappa_{850} \over 0.02\ {\rm cm}^{2}\,{\rm g}^{-1}}\right)^{-1}
\,M_\odot.
\label{e_cm}
\end{equation}
Anticipating the results below, the dust temperature is estimated
to be $T_d = 20\,$K. Thus, the measured masses in Orion A South 
range from $0.3\,M_\odot$ to $22\,M_\odot$. The mass versus size 
relation for the clumps is plotted in Figure \ref{f_oas_mr}.

\subsection{Modeling Clumps as Bonnor-Ebert Spheres}

In order to better estimate the clump properties, such as temperature and mass, 
and the environment in which the clumps are embedded, 
{it is useful} to equate the
observed clumps with simple physical constructions. 
The simplest static models 
{are spherical, isothermal, self-gravitating pressure-confined clumps
known as} Bonnor-Ebert spheres (Ebert 1955, Bonnor 1956, Hartmann 1998), a
one-dimensional family of equilibrium structures equating the importance of 
self-gravity, and possible collapse, to the degree of central concentration 
within the clump. In Paper II we developed a technique for identifying the 
internal clump temperature, the clump mass, and the confining
surface pressure on the clump, from the total flux at 850\,$\mu$m, $S_{850}$,
the observed size of the clump in arcseconds, ($R\arcsec_{\rm eff}$),
and the concentration, $C$, of the clump. Given the FWHM beamsize,
$B\arcsec$, of the observations, and the peak surface brightness,
${F}_{850}$ in Jy/beam, the concentration is defined as
\begin{equation}
C = 1 - { 1.13\,{B\arcsec}^2\,S_{850} \over
          \pi\,{R\arcsec_{\rm eff}}^2\,{F}_{850}}.
	  \end{equation}
For equilibrium Bonnor-Ebert spheres, the concentration
ranges from $C = 0.33$ for uniform density, non-self-gravitating objects
to $C = 0.72$ for critically self-gravitating objects. 
{\dij It is important to note that $C$ is an observationally determined quantity and
thus depends on both the resolution and sensitivity of the instrument used. Additionally,
the manner in which the objects are identified (Gaussian, wavelet, or {\tt clfind}) will
affect the measured concentration. Recent findings that the concentration correlates
with the existence of an embedded protostar (Walawender et al. 2005, 2006; Jorgensen et al.
2006) has focused interest on this quantity and comparison between the different object
identification techniques is being investigated.}  {For further
discussion on the complications of this procedure, refer to Papers II, III,
and IV.}

The concentration of each observed clump is plotted in Figure \ref{f_oas_con}
against both the clump mass, assuming $T_d = 20\,$K, and the clump size.
Five clumps appear more concentrated than allowed for by Bonnor-Ebert
spheres and are flagged by diamonds while three clumps are not concentrated
enough to be Bonnor-Ebert spheres and are flagged by crosses.  These latter
clumps all have relatively low peak brightnesses and lie near bright sources.
Thus, their size is likely somewhat confused.

Applying the Bonnor-Ebert sphere analysis used in Papers II-IV, the physical
conditions obtained for each clump are plotted in Figure \ref{f_oas_be}. 
In Paper IV we argue that the internal pressure of the clump is
composed of equal contributions from thermal and non-thermal motions
and that the opacity is higher than was used in Papers II and III.
Combined, these two changes produce almost identical values for
clump masses and bounding pressures as was found using the assumptions
given in Papers II and III but lower the derived temperatures by a
factor of approximately two.
Figure \ref{f_oas_be} plots the external pressure versus 
the internal temperature for the clumps identified in this paper.
We find that the clump temperatures in Orion A are clustered around 
$T_d = 22 \pm 5\,$K. 
This result is very similar to the values obtained for Orion B South
(Paper IV) and Orion B North (Paper III - after accounting for changes 
in the values used for opacity and pressure support determination).
The external pressure required to confine the clumps is
$\log(k^{-1}\,P\,{\rm cm}^3\,{\rm K}^{-1}) = 6.0 \pm 0.2$, 
similar to that needed in Orion B 
(Papers III and IV), and a factor of three to ten lower than 
obtained for the much denser core of Ophiuchus. As discussed in
Paper III, this pressure lies within the range expected deep inside the
Orion cloud.

\subsection{Clump Temperatures Derived From Submillimeter Colors}

Comparison of the 850\,$\mu$m and 450\,$\mu$m SCUBA maps allows
for an independent measure of the dust temperature.
The dust temperature within each clump can be estimated from
the submillimeter spectral index, $\gamma$, where 
$S(\nu) \propto \nu^{\gamma}$. Assuming that the dust emission
is approximated by a radiating blackbody at temperature $T_d$ and
with an opacity $\kappa(\nu)$ which follows a power-law such
that $\kappa(\nu) \propto \nu^{\beta}$, then
\begin{equation}
S(\nu) \propto \nu^{\beta}\,B_{\nu}(T_d),
\end{equation}
which can be inverted to find $T_d$ from observations of $S(\nu)$.
Unfortunately, this technique requires knowledge of $\beta$ which has 
been shown to vary significantly with environment (Goldsmith, Bergin, 
\& Lis 1997; Visser et al. 1998; Hogerheijde \& Sandell 2000; 
Beuther, Schilke, \& Wyrowski 2004). Friesen et al.\ (2005) have
shown that it is possible to use a Fourier Transform Spectrometer 
at submillimeter wavelengths to measure $\beta$ directly, however,
present instruments only allow for observations of extremely bright 
sources.

An additional complication for utilizing multi-wavelength observations to 
estimate the dust temperature is that the beam size and shape can
vary significantly. This is especially acute when using 850\,$\mu$m 
and 450\,$\mu$m observations taken at the JCMT (Hogerheijde \& Sandell 2000). 
The 450\,$\mu$m beam is
poorly approximated by a single Gaussian component, with almost
half the flux associated with a broad component about $30\arcsec$ in
extent.  Thus, simple attempts to convolve the 450\,$\mu$m map with a 
single Gaussian to obtain the lower spatial resolution of the 850\,$\mu$m 
map tends to overestimate the contribution of 450\,$\mu$m emission on 
large spatial scales. Reid and Wilson (2005) propose a more complicated
but accurate approach in which both the maps are convolved with the complex
telescope beam profile of the other wavelength.
Here we simplify the technique somewhat by taking the typical beam 
profile at each wavelength, as was performed in Paper IV. Thus, the 
JCMT 450\,$\mu$m beam is modeled as two Gaussians with 8.5\arcsec and 
30\arcsec\ FWHM and relative peak intensities of 0.95 and 0.05 while the 
850\,$\mu$m beam is modeled as two Gaussians with 14.5\arcsec and 
30\arcsec\ FWHM and relative peak intensities of 0.95 and 0.05.
{The fluxes measured and tabulated in Table 2 are derived after 
convolution and using exactly the same pixels to define each clump in 
both maps.}

Assuming that $\beta = 2$, the dust temperature at the center of each
clump is estimated using the total flux in the convolved 850\,$\mu$m and 
450\,$\mu$m maps.  For those sources where the flux ratio
produces unphysical temperatures, a  value of $50\,$K is assumed.
Unphysical values are caused by the 450\,$\mu$m to 850\,$\mu$m ratio
becoming larger than a critical value and suggest that either 
$\beta > 2$ or the 450\,$\mu$m data is corrupted by contamination 
by nearby sources or poor flux calibration.  We note 
that the typical uncertainty in the flux calibration is 
10\% at 850\,$\mu$m and 25\% at 450\,$\mu$m (Jenness et al. 2002)
{which leads to a large uncertainty in the flux ratio and hence 
the derived submillimeter temperature}.
The typical dust temperature derived from the submillimeter color 
measurement is $T_d = 18 \pm 9\,$K, excluding
the sources with unphysical values, in reasonable agreement with the
Bonnor-Ebert temperatures. Table \ref{t_450850} catalogues the
derived submillimeter spectral index derived temperatures along with 
the Bonnor-Ebert values. A graphical comparison is presented in Figure
\ref{f_obs_temp}. There is no clear trend between the two temperature
derivations, rather the temperatures appear to be scattered randomly
within the bounds of the uncertainty in each measure. This result
is consistent with the analysis of Orion B South (Paper IV).

\subsection{The Clump Mass Distribution Function}

The distribution of clump masses is most easily investigated by plotting
the cumulative number $N(M)$ of clumps with masses greater than $M$.
Figure \ref{f_oas_cum} presents the observed clump cumulative mass function
for both a constant flux-to-mass ratio ($T = 20\,$K; thin line) and for
masses calculated using the Bonnor-Ebert sphere analysis (thick line).
{\dij To represent the data with simple power-law fits, $N(M) \propto M^{-\alpha}$,
requires $\alpha \sim 2.0 \pm 0.5$} for moderate mass clumps ($M > 3 M_\odot$), 
while the lower mass clumps are well represented by $\alpha \sim 0.5$ 
(the Salpeter IMF has a slope of $\alpha = 1.35$ in this representation). 
We stress that incompleteness may be important at the low mass end, 
biasing $\alpha$ to lower values (c.f. Figure \ref{f_oas_mr} and 
discussion in Paper II). 

The results of this analysis are somewhat 
inconsistent with those in Papers II-IV
as well as the work of other submillimeter data sets obtained in
nearby molecular clouds (Motte et al. 1998; Testi \& Sargent 1998; 
Motte et al. 2001).  Typically, the continuum clump mass distribution has 
been comparable to the stellar Initial Mass Function (Scalo 1986; Kroupa 
et al. 1993; Kroupa 2002), 
at least for the more massive clumps observed. In Orion A South, 
however, the clump mass distribution {appears} {\it steeper} than the 
stellar Initial Mass Function, {similar to the results of Reid \& Wilson
(2005)}.  Note that all the observed submillimeter 
continuum clump mass distributions are significantly steeper, at least at the 
high mass end, than the distribution of mass inside molecular clouds obtained 
from CO measurements (Williams \& McKee 1997; McKee \& Williams 1997; Kramer 
et al. 1998; Williams, Blitz, \& McKee 2000 and references therein).

It is possible that the steep mass function of clumps is related to the
steep mass function of stars. In both cases, the majority of mass is
accounted for by small objects, unlike the molecular cloud distribution where
the rare massive clouds account for the majority of the mass.
Perhaps then the mass function of stars is determined before  collapse 
occurs. Only a small fraction, of order a few percent, of the cloud mass 
is observed to be in measured submillimeter clumps (Johnstone, Di Francesco,
\& Kirk 2004; Kirk, Johnstone, Di Francesco 2006), 
roughly consistent with the overall star formation 
efficiency in clouds. In contrast, the molecular gas distribution inside
clouds, observed in CO, contains the bulk of the cloud mass and is
therefore tracing the {large-scale} complexity of the cloud. When higher
density tracers are observed the gas mass fraction is found to be
similar to that of the submillimeter clumps, for example H$^{13}$CO 
line emission (Onishi et al. 2002). As well, {some} observations
of more distant molecular clouds detect larger, more massive, submillimeter
clumps that are observed to have a distribution similar to the CO observations
(Kerton et al. 2001; Mookerjea et al. 2004).  These results suggest that the 
there is a significant change in the structure of a molecular cloud on small,
dense, scales and that the process which drives the formation of these 
entities is different from the mechanism which produces the large scale 
structure in clouds.

\section{Clump Locations and Comparison with H$_2$}

The mapped portion of Orion A contains several active centers of star
formation which are spawning small clusters of stars, including
OMC4, OMC5, L1641N, HH34, and HH1/2.  In order to understand the 
connection between the submillimeter clumps and the formation of stars,
it is useful to correlate the SCUBA clumps with excited H$_2$ 2.122$\mu$m
emission which traces the energetic shocks known to emerge from young protostars. 
Fortunately, Stanke (2000) and Stanke, McCaughrean, and Zinnecker (2002;
SMZ) mapped most of the Orion A cloud in this
transition, providing an excellent resource for this analysis.

\subsection{Correlation of Clumps with H$_2$ Shock Emission}

The presence of shock excited H$_2$ emission near individual clumps provides
an attractive sign post for an embedded protostellar source. Walawender
et al. (2005, 2006) have shown that the combination of 2.122\,$\mu$m H$_2$
shock observations and 850\,$\mu$m continuum emission maps can help 
identify which submillimeter clumps are associated with Class 0 sources. 
Walawender et al. find that the presence of H$_2$ emission correlates
reasonably well with the concentration of the submillimeter clumps, such that
more concentrated sources have a higher likelihood of harboring embedded
protostars. This is to be expected from the simple theoretical identification
of the clumps with Bonnor-Ebert spheres, where higher concentration 
corresponds to the clump being closer to gravitational instability.

The investigations by Walawender et al. (2005, 2006) of regions within 
Perseus provide statistics on  only a handful of submillimeter clumps.
The present analysis of Orion A South allows for a much more robust 
investigation. Seventy of the seventy-one clumps obtained from the
submillimeter map lie within the boundary of the H$_2$ 2.122\,$\mu$m 
region mapped by Stanke (2000) and seventeen of these clumps have clearly 
associated H$_2$ emission (see Table 3 and Figures 9 through 17).  
A comparison between the presence of 
H$_2$ and the observational (physical) properties of the submillimeter clumps
strengthens the results of Walawender et al.  Tables 4-6 tabulate 
the correlation between the presence of molecular hydrogen 
versus total flux (mass), peak flux (column density), and concentration 
(importance of self-gravity).  There exists no discernible correlation 
between the presence of H$_2$ emission and total flux,  suggesting that 
clump mass is not a significant determinant for collapse of a starless core. 
There are, however, strong correlations with peak flux and concentration,
suggesting that column density and the balance between self-gravity and
thermal stability are significant in determining collapse.

\subsection{Details on Individual Regions with Orion A South}

\subsubsection{OMC4}

The southern portion of the ISF contains a cluster of about two dozen
SCUBA sources embedded in a pair of V-shaped north-south ridges, denoted
OMC4 by Johnstone \& Bally (1999). 
The northern end of this cluster of clumps appears to be
interacting with the southern portion of the Orion Nebula. 
An extensive network of fluorescent H$_2$ filaments wrap around
the northern boundary of these clumps.  Additionally, high resolution 
Hubble Space Telescope (HST) H$\alpha$ images reveal many young stars, 
proplyds, and Herbig-Haro objects in this region (Bally et al. 2006).  
There is not a simple correspondence between individual YSOs and SCUBA 
clumps, suggesting that little circumstellar material remains for
the optically visible sources. It is likely 
that the expanding Orion Nebula has recently overrun this region, uncovering 
many young stars.  Interestingly, few SCUBA clumps are coincident with IRAS point 
sources, 2 $\mu$m sources, or H$_2$ outflows.  Thus, most of these clumps 
do not contain obvious protostars or YSOs.  Stanke (2000) does report several 
H$_2$ flows in the region (Stanke 2-8, 2-9, 3-1, 3-4, 3-5, 3-8, and
3-10).  These may originate in more mature YSOs which have shed their
dusty submillimeter envelopes.  Figure 9 shows the OMC4 region.   

A large, arcminute-scale H$_2$ flow consisting of Stanke 2-7 and 3-2 
(SMZ29) emerges from a dim SCUBA clump located at J(2000) =  5:34:40.9, -5:31:47
about 8\arcmin\ west of the OMC4 region at PA $\sim$ 340\arcdeg .  
The clump is clearly real, but located too close to the edge of the 
mapped SCUBA field for inclusion in our clump catalog.
A bright K-band reflection nebula opens towards the north-northwest.

Stanke 3-11 is a northeast facing bow shock emerging from SMM 053474-05414 
(ID 01) at PA $\sim$ 40\arcdeg .  
IRAS 05322-0543 may be associated with this clump and outflow.

The K-band star at J(2000) = 5:34:35.7, -5:40:10 embedded in
a dim, uncatalogued SCUBA clump located at the western edge of our 
mapped region may be the driving source of the chain of H$_2$ shocks 
Stanke 3-6, 3-7, and 3-9 (SMZ31) which extends to the northwest at 
PA $\sim$ 335\arcdeg with respect to this clump.

Finally, a fan-shaped reflection nebula opens towards the northeast
from another uncatalogued SCUBA clump at J(2000) = 5:34:48.9, -5:41:42
at the southern end of OMC4.  No shocks are detected along 
its axis.

\subsubsection{OMC5 behind NGC 1980}

The region situated at the southern end of the ISF about 30\arcmin\
south of the Orion Nebula contains a loose cluster of SCUBA clumps 
located immediately behind the star cluster NGC 1980.  We denote
the region as OMC5. The cluster of submillimeter clumps contains at 
least one IRAS source (IRAS 05328-6000), but this is a lower limit 
to embedded YSOs as confusion within this region limits the usefulness 
of the IRAS data.

The faint H$_2$ jet, Stanke 3-17 (SMZ37), emerges parallel to a 
reflection nebula at PA $\sim$ 35\arcdeg). This feature emerges from
the southern edge of a faint SCUBA clump (not listed in Table 1)
at J(2000) = 5:35:04.5, -5:51:50.  Stanke 3-16 north of this feature
appears to emerge from SMM 053514-05513 (ID 25) at PA $\sim$ 320\arcdeg .
Figure 10 shows the H$_2$ emission and 850 $\mu$m contours in this region. 

A chain of H$_2$ shocks was detected by Stanke (2000) which consists of
Stanke 4-1, 4-2, and 4-3 (Position Angle, PA $\sim$ 335\arcdeg) 
emerging from a faint, fan-shaped reflection nebula oriented towards 
the north-northwest.  These features appear to emanate from SCUBA clump 
SMM 053515-05558 (ID 30) which exhibits a faint, K-band reflection nebula opening 
towards the north-northwest (see Figure 11).  
The morphology of these latter shocks 
consist of north-facing bows consistent with an outflow emerging 
from the clump.   Stanke 4-5 may trace a counterflow on the
other side of this clump, or possibly a flow from an adjacent clump.  
This sub-millimeter clump apparently contains a Class 0 YSO driving a 
bipolar outflow visible in H$_2$ (SMZ38).  No optical HH objects are 
detected in the vicinity of this flow, therefore it must be embedded.  

The H$_2$ features Stanke 4-4, 4-5, 4-6 and 4-7 cluster 
around SMM 053520-05571 (ID 37) and may trace outflows emerging from this 
clump., including SMZ39.  SCUBA clump SMM 053515-05584 (ID 31) 
south of SMZ38 is at the center of the flow SMZ39
and therefore may contain the driving source.

Several isolated SCUBA clumps are located south of 
NGC 1980.  Three infrared point sources are embedded within 
this region.  IRAS 05331-0606 is located at the north-end of a ridge 
containing the diffuse dust clumps SMM 053568-06057 (ID 44), 
053571-06077 (ID 45), and 053573-06078 (ID 46).  
A bright H$_2$ shock and HH object located at
J(2000) =  5:35:37.3, -6:02:32 several arcminutes north-northeast of
IRAS 05331-0606 has a bow shape consistent with a flow emerging from
the vicinity of this IRAS source (Stanke 4-9; SMZ40 - not shown).

IRAS 05334-0611 is associated with a star embedded within SMM 
053586-06100 (ID 47).  This star appears to drive a pair of H$_2$ bubbles
designated Stanke 5-2 (SMZ43) towards PA $\sim$ 70/250\arcdeg 
(Figure 12).

IRAS 05329-0614 is coincident with SMM 053537-06131 (ID 40) and 
appears to drive a long 17\arcmin\ long H$_2$ and 
HH flow towards PA $\sim$ 70/250\arcdeg\ (see Figure 12).  
A conical reflection 
nebula opening towards the southwest is evident in the K-band images of 
Stanke (2000) and a 10 to 20\arcsec\ long H$_2$ jet (Stanke 4-12) emerges 
along this axis.   A large (3\arcmin\ diameter), faint, complex 
of HH objects seen as faint [\sii ] and \Ha\ filaments are located 
14\arcmin\ southwest of this source between J(2000) = 5:34:22.8, 
-6:18:19 and 5:34:24.22   -6:15:13.  A bright but compact 
HH object is found 3\arcmin\ to the northeast at J(2000) = 
5:35:34.75   -6:12:04.  The curvatures of the various filaments
in these HH objects are consistent with being driven by a jet or
outflow from IRAS 05329-0614. The H$_2$  portions of this flow are
designated Stanke 4-11 and 4-12 (SMZ42).

SCUBA clump SMM 053660-06149 (ID 65), located due north of the L1641N cluster, 
contains a K-band source and a wide-angle H$_2$ flow 
(Stanke 5-5; SMZ46) that appears to propagate towards 
PA $\sim$ 20/200\arcdeg .  The clump is located at the southeast end 
of a filament of molecular gas and chain of clumps
(SMM 053618-06106 (ID 48), 053637-06129 (ID 55), 053641-06140 (ID 57), 
and 053645-06145 (ID 61)) which contain no IRAS sources or H$_2$ objects.

\subsubsection{L1641N}

This region contains a dense cluster of about 100 embedded stars, including
moderate-luminosity IR sources, including IRAS 05338-0624 and IRAS 05339-0626
(Figure 13).  
Dozens of HH and H$_2$ flows (Stanke Field 5) appear to be bursting out 
of this cluster in all directions (Stanke 2000; Reipurth, Bally, and 
Devine 1997).  At least two flows have reached parsec-scale dimensions, and 
their northern lobes have propagated large bow shocks currently impinging 
on the outskirts of the Orion Nebula about a degree to the north.
About a dozen SCUBA clumps are associated with the L1641 cluster.
The region is too confused to make specific associations between 
individual outflows, clumps, and stars (see Figure 13). 
The outflows SMZ48, 49, 50, 51, 53, and 54 emerge from this cluster.
Reipurth et al. (1997) proposed that several giant HH flows extend for
beyond the molecular cloud for over 10 pc from this cluster.  One of these
flows may be related to the north-south H$_2$ flow SMZ 49.   It has also been
suggested that the giant HH bow shock HH131 located 2.5 degrees south originates
here (Wang et al. 2005).

SCUBA clump SMM 053641-06226 (ID 58) is located on the east side of the L1641N
cluster and appears to be associated with a large H$_2$ and Herbig-Haro
flow SMZ51 that emerges from the L1641N region towards PA $\sim$ 80\arcdeg\
and that is associated with HH 301 and possibly HH 302.

The flow SMZ54 emerges from a prominent disk shadow in both visual and
near-IR images located midway between two SCUBA clumps (ID 54 and 58).  
Although we did not resolve a clump associated with this edge-on disk, 
there is bright extended 850 $\mu$m emission in this region.

SCUBA clump SMM053631-06221 (ID 52) is embedded in the center of the L1641N cluster
and may be associated with IRAS 05338-0624.  This object is located at the
base of the giant SMZ 49 outflow which is associated with HH 306 -- 310.
The orientation and shapes of associated bow shocks indicate
that this flow is propagating at PA $\sim$ 355\arcdeg\ through a
giant cavity on the east side of the Integral Shaped Filament.
The terminal shocks of this flow at its north end may be
interacting with the southern boundary of the Orion Nebula
(Reipurth et al. 1998).

\subsubsection{HH 34}

The SCUBA clump 053550-06269 (ID 41) is associated with the spectacular HH 34 jet and 
the chain of HH objects which delineate its parsec-scale outflow
(Devine et al. 1997; SMZ55 - indicated by a dashed line in Figures 14 and 15).  
Another clump, 053551-06265 (ID 42), appears associated with
HH 34 IRS5, an embedded near-IR star responsible for a bright
south-facing reflection nebula located on the eastern edge of the 
HH 34 cloud core.  The H$_2$ flow Stanke 5-24 (SMZ56) appears
to trace an embedded jet from yet another star in this region,
although no connection with a SCUBA clump is observed. Thus,
the cloud core surrounding the HH~34 source appears 
to have spawned a small group of YSOs.   

A diffuse group of faint SCUBA clumps lies half-way between HH 34 and
L1641N several arcminutes southeast of the enigmatic object HH 222
whose filaments resemble a waterfall (Castets et al. 2004).   
At least one K-band star in the complex of
faint dust emission illuminates a fan-shaped reflection nebula opening
towards the southwest.  It is possible that HH~222 wraps around
the northwest rim of this cluster of dust clouds.  

\subsubsection{South of L1641N}

Several SCUBA sources south of L1641N but north of HH~1/2 drive
interesting outflows.  SCUBA clump SMM 053660-06389 (ID 66 - Figure 16) 
is not associated with an IRAS point source but drives a stunning 
bipolar H$_2$ flow at PA $\sim$ 170/350\arcdeg\ (Stanke 5-28 = SMZ59).  
The embedded source YSO is likely to be an isolated Class 0.  
The SCUBA clump is elongated along the 
outflow axis.  This source was mapped in CO by Levreault (1988)
and Morgan et al. (1991) who designated it as as V380 Ori NE.
Detailed H$_2$, interferometric CO, and SCUBA observations were
presented by Davis et al. (2000).

SCUBA source SMM 053700-06272 (ID 70 - Figure 16), located several 
arcminutes northeast of SMM 053660-06389, is associated with a northeast 
facing reflection nebula and drives an H$_2$ flow, Stanke 5-27 (SMZ58).
A faint [\sii ] dominated  HH knot is associated with the northeast end 
of this flow.  The embedded YSO is likely to be a Class O source since there is 
no IRAS source coincident with this clump.
The SCUBA clump is elongated orthogonal to the outflow axis.

The several IRAS sources in this region are not associated with
SCUBA clumps.  For example, IRAS 05342-0635 seems to be associated with
a visible and near-IR star which exhibits no outflow activity
in either H$_2$ or visual wavelength images.  The dust from this star
may trace a remnant disk or envelope which has not fueled accretion
onto the star in recent times. In contrast,  IRAS 05345-0635 is
associated with the $m_R$ = 12 magnitude star to BE Ori which drives a 
spectacular HH jet at PA $\sim$ 45\arcdeg\ (SMZ57).   Interestingly,
this star is not associated with a SCUBA source.  Perhaps it lacks an
extended, cool, and inactive outer envelope but still retains an inner disk
which is actively fueling accretion and outflow.

\subsubsection{NGC~1999 and HH 1/2}

A cluster of 4 SCUBA clumps, 053629-06474 (ID 50) , 053631-06455 (ID 53), 
053638-06461 (ID 56), and 053642-06249 (ID 60), are located south
of the visually-bright NGC 1999 reflection nebula powered by V380 Ori
and are associated with YSOs in the vicinity of the Herbig-Haro 
objects HH~1/2 (see Figure 17).  Interestingly, NGC 1999/V380 Ori, at 
the northern end of this sub-cluster, is not coincident with a SCUBA source.  Thus, 
this star must be relatively clear of circumstellar dust.  

Clump SMM 053638-06461 (ID 56), centered on the source of the HH 1/2 bipolar outflow, is 
the brightest 850 $\mu$m source in this complex (see Figure 17).  
IRAS did not detect a point
source here; nevertheless, the YSO embedded in this clump drives
one of the brightest and most spectacular HH flows in the sky, the
objects HH 1 and 2 (Stanke 6-10 and 6-15; SMZ64). 

The second brightest clump, SMM 053631-06455 (ID 53),
located near the H$_2$O maser source 
northwest of HH 1/2 is associated with
IRAS 05338-0647.  Stanke 6-11 is located near this source.  From the
morphology of the H$_2$ emission, it is likely that this feature and
Stanke 6-6 (which coincides with HH 3) are driven by this source. 

Clump SMM 053641-06447 (ID 59), located between the HH 1/2 source region and NGC 1999
is associated with IRAS 05339-0646, a visible star ($m_R$ $\sim$ 16 mag
L1641-KMS31), several faint HH objects (HH 147), and the H$_2$ features 
Stanke 6-12 (SMZ63).

Clump 053629-06474 (ID 50), located several 
arcminutes southwest of HH 1/2 source, is 
the brightest part of several filaments of dust emission which point 
radially away from clump 053642-06249, located northeast of the the HH 1/2 core.
No IRAS source, star or outflow is associated with this feature.  
These radial filaments are  reminiscent of the network of
dust filaments which point away from luminous centers of star formation
in the OMC1 cloud core in the middle of the ISF, and from the NGC 1333
star forming complex in the Perseus molecular cloud.
The filaments may be the walls of ancient outflow cavities.
Alternatively they may be cometary cloud-tails located in the 
shadows of dense clumps illuminated by UV or shock-heated by winds.

\section{Conclusions}\label{s_sum}

Using the techniques developed in Papers I-IV we have identified 71 
submillimeter clumps within a  2300 arcmin$^2$ region of Orion A South.
We have been able to derive estimated masses and temperatures for the
majority of the clumps by noting that they can be modeled as
Bonnor-Ebert spheres; that is, they can be approximated by almost
constant density models with low internal velocities. On this basis
the typical clump temperature is found to be 22$\pm$5~K; this value is
in reasonable agreement with that (18$\pm$9~K) derived from the
spectral indices obtained from the 850 and 450\,$\mu$m data.  Thus for
most of the clumps a temperature of 20~K is a good working
approximation, similar to the results in Orion B (Papers III and IV).  

The mass function of submillimeter clumps found in Orion A South is
steeper than the molecular cloud mass spectrum, suggesting that
most of the  mass of submillimeter clumps lies in small mass objects.
This is in agreement with the results in other nearby regions
(Motte et al. 1998; Paper II; Paper III; Motte et al. 2001; Paper IV) although
the mass function found here is somewhat steeper still than the
mass functions derived for these other regions.

A comparison of the appearance of molecular hydrogen shock excited
emission and the location of submillimeter clumps reveals that
concentration and central column density of the clumps are correlated
with evidence of internal protostellar activity. Several new Class
O sources have been identified in Orion A South using this technique.

\acknowledgments

The research of D.J.\ is supported through a grant
from the Natural Sciences and Engineering Research Council of Canada.  
The JCMT is operated by the Joint Astronomy Centre on behalf of the Particle 
Physics and Astronomy Research Council of the UK, the Netherlands Organization 
for Scientific Research, and the National Research Council of Canada.
The authors acknowledge the data analysis facilities provided by the 
Starlink Project which is run by CCLRC on behalf of PPARC. We thank
the anonymous referee for suggestions which significantly improved this paper.
We thank Thomas
Stanke for providing us with his 2.122$\mu$m images in electronic form. We
also thank Rachel Friesen and Helen Kirk for critical reading of this manuscript.
J.B. acknowledges support from NASA grants NAG5-8108 (LTSA),
the University of Colorado Center for Astrobiology and
NASA grant NCC2-1052, and National Science Foundation grants
AST 0407356 and AST 9819820.

\newpage


\clearpage





\begin{deluxetable}{lcccrrclrcrr}
\tabletypesize{\scriptsize}
\tablewidth{480pt}
\tablecaption{Clump properties in Orion A South derived from  850\,$\mu$m data.\label{t_850data}}
\tablehead{
\colhead{Name\tablenotemark{a}}&
\colhead{ID}&
\colhead{R.A.\tablenotemark{b}}&
\colhead{Dec.\tablenotemark{b}}&
\colhead{$S_{850}$\tablenotemark{c}}&
\colhead{$S_{850}^{\rm peak}$\tablenotemark{c}}&
\colhead{$R_{\rm eff}$\tablenotemark{c}}&
\colhead{$C$\tablenotemark{d}}&
\colhead{$T_d$\tablenotemark{d}}&
\colhead{log$P/k$\tablenotemark{d}}&
\colhead{$M_{T_d}$\tablenotemark{d}}&
\colhead{$M_{T_d=20\,{\rm K}}$\tablenotemark{e}}\\
\colhead{(SMM J)}&
\colhead{}&
\colhead{(J2000)}&
\colhead{(J2000)}&
\colhead{(Jy)}&
\colhead{(Jy/bm)}&
\colhead{(10$^3$ AU)}&
\colhead{}&
\colhead{(K)}&
\colhead{(K/cm$^{3}$)}&
\colhead{($M_\odot$)}&
\colhead{($M_\odot$)}
}

\startdata
053474-05414&01&  05:34:44.2& -05:41:27&  0.78&  0.41&    7&   0.47& 14&    6.3&   1.11&   0.63\\
053484-05462&02&  05:34:50.6& -05:46:15&  2.81&  0.35&   13&   0.39& 25&    6.0&   1.64&   2.26\\
053486-05389&03&  05:34:51.4& -05:38:54&  0.76&  0.29&    7&   0.36& 24&    6.2&   0.47&   0.61\\
053487-05423&04&  05:34:52.0& -05:42:18&  1.73&  0.24&   12&   0.33& 29&    5.9&   0.83&   1.39\\
053490-05463&05&  05:34:54.0& -05:46:18&  4.23&  0.75&   13&   0.53& 20&    6.2&   3.40&   3.40\\
053492-05435&06&  05:34:55.2& -05:43:30&  7.12&  0.42&   20&   0.39& 31&    5.9&   3.12&   5.72\\
053494-05460&07&  05:34:56.2& -05:46:03&  7.76&  1.20&   17&   0.69& 20&    5.9&   6.23&   6.23\\
053495-05415&08&  05:34:57.0& -05:41:33&  6.46&  0.44&   19&   0.42& 27&    5.9&   3.40&   5.19\\
053496-05435&09&  05:34:57.6& -05:43:33&  2.24&  0.29&   13&   0.39& 23&    5.9&   1.47&   1.80\\
053497-05365&10&  05:34:58.2& -05:36:30&  2.17&  0.31&   12&   0.34& 32&    6.0&   0.91&   1.74\\
053497-05409&11&  05:34:58.2& -05:40:54&  4.43&  0.37&   17&   0.46& 21&    5.9&   3.31&   3.56\\
053499-05441&12&  05:34:59.4& -05:44:09&  1.96&  0.24&   13&   0.33& 30&    5.9&   0.90&   1.58\\
053500-05389&13&  05:35:00.2& -05:38:57&  6.87&  0.61&   17&   0.49& 23&    6.0&   4.51&   5.53\\
053501-05400&14&  05:35:00.4& -05:40:03&  2.70&  0.42&   13&   0.48& 18&    6.0&   2.55&   2.17\\
053503-05556&15&  05:35:01.8& -05:55:36&  3.64&  0.53&   14&   0.53& 18&    6.0&   3.43&   2.92\\
053504-05372&16&  05:35:02.6& -05:37:12&  4.12&  0.61&   12&   0.39& 30&    6.2&   1.89&   3.31\\
053504-05379&17&  05:35:02.4& -05:37:54&  6.79&  0.76&   15&   0.49& 24&    6.1&   4.20&   5.45\\
053505-05362&18&  05:35:03.0& -05:36:12& 11.03&  0.98&   19&   0.55& 24&    6.0&   6.82&   8.87\\
053508-05374&19&  05:35:05.0& -05:37:24&  6.68&  1.02&   13&   0.49& 25&    6.3&   3.91&   5.37\\
053509-05348&20&  05:35:05.2& -05:34:48&  9.60&  0.96&   17&   0.54& 24&    6.1&   5.94&   7.72\\
053510-05331&21&  05:35:06.0& -05:33:09&  3.70&  0.62&   14&   0.57& 17&    6.0&   3.81&   2.97\\
053511-05339&22&  05:35:06.6& -05:33:57&  3.62&  0.52&   13&   0.43& 23&    6.1&   2.38&   2.91\\
053513-05567&23&  05:35:07.6& -05:56:42&  1.01&  0.37&    8&   0.44& 16&    6.2&   1.15&   0.81\\
053514-05359&24&  05:35:08.4& -05:35:57&  9.95&  1.38&   16&   0.60& 24&    6.2&   6.15&   8.00\\
053514-05513&25&  05:35:08.4& -05:51:21&  4.69&  0.62&   15&   0.51& 20&    6.0&   3.77&   3.77\\
053514-05579&26&  05:35:08.6& -05:57:57&  1.85&  0.38&   10&   0.39& 23&    6.2&   1.22&   1.49\\
053514-06136&27&  05:35:08.4& -06:13:39&  4.45&  0.45&   16&   0.48& 20&    5.9&   3.58&   3.58\\
053515-05342&28&  05:35:08.8& -05:34:15&  3.84&  0.38&   14&   0.28& 39&    6.1&   1.26&   3.09\\
053515-05533&29&  05:35:08.8& -05:53:18&  0.83&  0.30&    8&   0.41& 17&    6.2&   0.86&   0.67\\
053515-05558&30&  05:35:09.0& -05:55:51&  7.67&  1.50&   17&   0.75& 19&    5.8&   6.66&   6.17\\
053515-05584&31&  05:35:09.2& -05:58:27&  2.50&  0.69&   12&   0.65& 15&    6.0&   3.16&   2.01\\
053516-05379&32&  05:35:09.8& -05:37:54&  1.64&  0.37&   10&   0.43& 18&    6.1&   1.54&   1.32\\
053516-05522&33&  05:35:09.4& -05:52:12&  2.92&  0.52&   12&   0.48& 19&    6.1&   2.53&   2.34\\
053517-05351&34&  05:35:10.4& -05:35:06&  6.47&  0.94&   13&   0.48& 25&    6.3&   3.78&   5.20\\
053518-06139&35&  05:35:10.8& -06:13:57&  8.40&  0.54&   21&   0.54& 21&    5.8&   6.28&   6.75\\
053520-05345&36&  05:35:12.0& -05:34:30&  6.35&  0.94&   15&   0.62& 20&    6.0&   5.10&   5.10\\
053520-05571&37&  05:35:12.2& -05:57:06&  2.20&  0.48&   11&   0.47& 18&    6.1&   2.07&   1.77\\
053523-05579&38&  05:35:13.6& -05:57:57&  4.19&  0.85&   14&   0.66& 17&    5.9&   4.31&   3.36\\
053524-05332&39&  05:35:14.6& -05:33:15&  1.73&  0.26&   12&   0.37& 25&    6.0&   1.01&   1.39\\
053537-06131&40&  05:35:22.0& -06:13:06&  1.24&  0.29&    9&   0.36& 27&    6.1&   0.65&   0.99\\
053550-06269&41&  05:35:29.8& -06:26:57&  3.40&  0.94&   14&   0.76& 15&    5.8&   4.30&   2.74\\
053551-06265&42&  05:35:30.6& -06:26:30&  1.89&  0.39&   11&   0.45& 18&    6.1&   1.78&   1.52\\
053562-06097&43&  05:35:37.2& -06:09:42&  0.39&  0.30&    5&   0.37& 20&    6.4&   0.32&   0.32\\
053568-06057&44&  05:35:41.0& -06:05:45&  2.36&  0.24&   14&   0.36& 31&    5.8&   1.04&   1.90\\
053571-06077&45&  05:35:42.4& -06:07:45&  1.98&  0.28&   12&   0.35& 31&    6.0&   0.87&   1.59\\
053573-06078&46&  05:35:43.6& -06:07:48&  2.10&  0.28&   12&   0.34& 31&    6.0&   0.92&   1.69\\
053586-06100&47&  05:35:51.8& -06:10:00&  1.84&  0.57&   10&   0.60& 14&    6.0&   2.62&   1.48\\
053618-06106&48&  05:36:10.8& -06:10:39&  5.32&  0.87&   16&   0.68& 17&    5.8&   5.49&   4.28\\
053628-06240&49&  05:36:17.0& -06:24:00&  1.70&  0.30&   11&   0.39& 22&    6.0&   1.19&   1.37\\
053629-06474&50&  05:36:17.6& -06:47:27&  1.47&  0.24&   11&   0.35& 28&    6.0&   0.74&   1.18\\
053630-06233&51&  05:36:18.2& -06:23:21&  3.63&  0.40&   15&   0.44& 22&    6.0&   2.54&   2.92\\
053631-06221&52&  05:36:18.6& -06:22:09& 27.48&  4.69&   24&   0.86& 29&    5.9&  13.14&  22.09\\
053631-06455&53&  05:36:18.4& -06:45:30&  8.55&  1.31&   17&   0.69& 20&    5.9&   6.87&   6.87\\
053636-06234&54&  05:36:21.6& -06:23:24&  5.03&  0.46&   17&   0.46& 22&    5.9&   3.52&   4.05\\
053637-06129&55&  05:36:22.0& -06:12:54&  3.10&  0.30&   16&   0.43& 20&    5.8&   2.49&   2.49\\
053638-06461&56&  05:36:22.6& -06:46:09& 12.94&  1.99&   21&   0.79& 22&    5.8&   9.05&  10.40\\
053641-06140&57&  05:36:24.6& -06:14:00&  5.17&  0.57&   15&   0.48& 22&    6.0&   3.62&   4.16\\
053641-06226&58&  05:36:24.6& -06:22:36& 10.47&  0.88&   21&   0.64& 21&    5.8&   7.83&   8.41\\
053641-06447&59&  05:36:24.8& -06:44:42&  5.17&  0.65&   17&   0.60& 18&    5.9&   4.88&   4.16\\
053642-06249&60&  05:36:25.0& -06:24:54& 14.00&  1.24&   21&   0.64& 24&    5.9&   8.66&  11.25\\
053645-06145&61&  05:36:27.2& -06:14:30&  3.82&  0.28&   17&   0.35& 36&    5.8&   1.38&   3.07\\
053648-06250&62&  05:36:28.8& -06:25:00&  2.02&  0.28&   12&   0.31& 31&    6.0&   0.89&   1.63\\
053652-06252&63&  05:36:31.2& -06:25:15&  1.89&  0.33&   11&   0.41& 21&    6.1&   1.42&   1.52\\
053655-06262&64&  05:36:32.8& -06:26:15&  3.32&  0.34&   15&   0.43& 21&    5.9&   2.49&   2.67\\
053660-06149&65&  05:36:36.2& -06:14:57&  1.30&  0.38&    9&   0.50& 15&    6.1&   1.65&   1.05\\
053660-06389&66&  05:36:36.0& -06:38:54&  6.82&  1.09&   23&   0.84& 16&    5.4&   7.74&   5.48\\
053669-06262&67&  05:36:41.4& -06:26:15& 12.04&  0.90&   24&   0.67& 21&    5.7&   9.01&   9.68\\
053682-06287&68&  05:36:49.0& -06:28:42&  4.62&  0.36&   18&   0.48& 19&    5.8&   4.01&   3.72\\
053685-06298&69&  05:36:51.2& -06:29:51&  1.55&  0.25&   11&   0.35& 28&    5.9&   0.78&   1.25\\
053700-06372&70&  05:37:00.0& -06:37:12&  3.41&  0.61&   13&   0.58& 17&    6.0&   3.52&   2.74\\
053728-06498&71&  05:37:16.8& -06:49:48&  1.34&  0.49&    9&   0.53& 15&    6.2&   1.70&   1.08\\
\enddata
\tablenotetext{a}{Name formed from J2000 positions (hhmm.mmdddmm.m). }
\tablenotetext{b}{Position of peak surface brightness within clump (accurate to 3\arcsec).}
\tablenotetext{c}{Radius, peak flux, and total flux are derived from {\tt clfind} (Williams et al. 1994). The peak flux and total flux have uncertainties of about 20\%, mostly due to absolute flux calibration. The radius has not been deconvolved from the telescope beam.}
\tablenotetext{d}{Quantities derived from Bonnor-Ebert analysis (see text).}
\tablenotetext{e}{Mass derived from the total flux assuming $T_d = 20\,$K and
$\kappa_{850} = 0.02\,$cm$^{2}$g$^{-1}$.}
\end{deluxetable}





\begin{deluxetable}{lccrrc}
\tabletypesize{\scriptsize}
\tablewidth{400pt}
\tablecaption{Submillimeter wavelength properties of clumps in Orion A South \label{t_450850}}
\tablehead{
\colhead{Name\tablenotemark{a}}&
\colhead{$S_{450}/S_{850}$\tablenotemark{b}}&
\colhead{$S_{450}^{\rm peak}/S_{850}^{\rm peak}$\tablenotemark{b}}&
\colhead{$T_d$(BE)\tablenotemark{c}}&
\colhead{$T_d$(SM) Best\tablenotemark{d}}&
\colhead{$T_d$(SM) Range\tablenotemark{d}}\\
\colhead{(SMM J)}&
\colhead{}&
\colhead{}&
\colhead{(K)}&
\colhead{(K)}&
\colhead{(K)}\\
}

\startdata
053474-05414&    5.9&    6.0& 14& 13&  8-- 24\\
053484-05462&    6.4&    6.7& 25& 14&  8-- 31\\
053486-05389&    6.4&    6.9& 24& 14&  8-- 31\\
053487-05423&    6.1&    6.9& 29& 13&  8-- 26\\
053490-05463&    7.1&    7.1& 20& 16&  8-- 47\\
053492-05435&    6.7&    7.1& 31& 15&  8-- 35\\
053494-05460&    7.1&    6.7& 20& 16&  8-- 48\\
053495-05415&    7.6&    7.8& 27& 18&  9-- $\infty$\\
053496-05435&    7.3&    7.6& 23& 17&  9-- $\infty$\\
053497-05365&    8.0&    9.6& 32& 20&  9-- $\infty$\\
053497-05409&    7.2&    7.9& 21& 16&  9-- $\infty$\\
053499-05441&    5.7&    6.0& 30& 12&  7-- 23\\
053500-05389&    8.2&    9.3& 23& 21&  9-- $\infty$\\
053501-05400&    7.2&    8.1& 18& 16&  9-- $\infty$\\
053503-05556&    4.6&    4.9& 18& 10&  7-- 16\\
053504-05372&    9.8&   10.2& 30& 33& 11-- $\infty$\\
053504-05379&    9.1&    9.5& 24& 26& 10-- $\infty$\\
053505-05362&   10.8&   11.2& 24& $\infty$& 12-- $\infty$\\
053508-05374&    9.0&    9.6& 25& 25& 10-- $\infty$\\
053509-05348&   10.3&   10.7& 24& 40& 11-- $\infty$\\
053510-05331&    8.5&    9.6& 17& 22& 10-- $\infty$\\
053511-05339&    8.4&    9.7& 23& 22& 10-- $\infty$\\
053513-05567&    3.4&    3.5& 16&  8&  6-- 11\\
053514-05359&   10.2&   10.0& 24& 38& 11-- $\infty$\\
053514-05513&    4.7&    5.3& 20& 10&  7-- 16\\
053514-05579&    3.7&    4.8& 23&  9&  6-- 12\\
053514-06136&    6.7&    7.7& 20& 15&  8-- 37\\
053515-05342&   10.0&   11.4& 39& 35& 11-- $\infty$\\
053515-05533&    4.5&    4.5& 17& 10&  7-- 15\\
053515-05558&    4.2&    4.3& 19& 10&  6-- 14\\
053515-05584&    4.5&    4.6& 15& 10&  7-- 15\\
053516-05379&    4.3&    5.7& 18& 10&  6-- 14\\
053516-05522&    5.2&    5.1& 19& 11&  7-- 19\\
053517-05351&    9.8&    9.3& 25& 33& 11-- $\infty$\\
053518-06139&    6.8&    7.5& 21& 15&  8-- 38\\
053520-05345&   10.1&   10.5& 20& 37& 11-- $\infty$\\
053520-05571&    3.2&    3.5& 18&  8&  6-- 10\\
053523-05579&    3.3&    3.9& 17&  8&  6-- 11\\
053524-05332&   10.8&   11.7& 25& 49& 12-- $\infty$\\
053537-06131&    4.3&    5.0& 27& 10&  6-- 14\\
053550-06269&    2.0&    3.3& 15&  6&  5--  8\\
053551-06265&    3.3&    4.4& 18&  8&  6-- 11\\
053562-06097&    5.9&    6.0& 20& 13&  8-- 25\\
053568-06057&    7.7&    9.0& 31& 18&  9-- $\infty$\\
053571-06077&    8.0&    8.4& 31& 20&  9-- $\infty$\\
053573-06078&    7.4&    8.0& 31& 17&  9-- $\infty$\\
053586-06100&    9.5&    7.8& 14& 29& 10-- $\infty$\\
053618-06106&    5.1&    6.6& 17& 11&  7-- 18\\
053628-06240&    8.4&    9.3& 22& 21&  9-- $\infty$\\
053629-06474&    5.1&    5.3& 28& 11&  7-- 18\\
053630-06233&    7.8&   11.4& 22& 19&  9-- $\infty$\\
053631-06221&    8.4&    7.4& 29& 22&  9-- $\infty$\\
053631-06455&    8.2&    9.0& 20& 21&  9-- $\infty$\\
053636-06234&    7.7&    7.8& 22& 18&  9-- $\infty$\\
053637-06129&    7.8&    9.0& 20& 19&  9-- $\infty$\\
053638-06461&    8.4&    9.0& 22& 22&  9-- $\infty$\\
053641-06140&   12.0&   11.4& 22& $\infty$& 13-- $\infty$\\
053641-06226&    7.2&    8.0& 21& 16&  8-- 49\\
053641-06447&    7.1&    6.6& 18& 16&  8-- 46\\
053642-06249&    9.0&    8.4& 24& 25& 10-- $\infty$\\
053645-06145&   11.3&   13.5& 36& $\infty$& 12-- $\infty$\\
053648-06250&    6.6&    9.0& 31& 15&  8-- 34\\
053652-06252&    6.8&    7.0& 21& 15&  8-- 38\\
053655-06262&    6.5&    6.7& 21& 14&  8-- 32\\
053660-06149&    7.3&    9.8& 15& 17&  9-- $\infty$\\
053660-06389&    6.8&    7.1& 16& 15&  8-- 38\\
053669-06262&    8.3&    8.1& 21& 21&  9-- $\infty$\\
053682-06287&    8.7&    9.7& 19& 23& 10-- $\infty$\\
053685-06298&    9.6&    8.8& 28& 30& 11-- $\infty$\\
053700-06372&    6.6&    5.8& 17& 15&  8-- 34\\
053728-06498&    6.0&    6.7& 15& 13&  8-- 26\\
\enddata
\tablenotetext{a}{Name formed from J2000 positions (hhmm.mmdddmm.m). }
\tablenotetext{b}{The 850\,$\mu$m and 450\,$\mu$m peak flux and total flux are derived from {\it clfind} (Williams et al. 1994) after convolution to an identical beam size. The uncertainty in the flux ratios is approximately 50\%.}
\tablenotetext{c}{Quantity derived from Bonnor-Ebert analysis (see text).}
\tablenotetext{d}{Quantity derived from spectral energy fit to 450 and 850$\mu$m integrated fluxes (see text). The range of acceptable temperatures is determined assuming a 50\% uncertainty in the flux ratio measurement.}
\end{deluxetable}





\begin{deluxetable}{lcl}
\tabletypesize{\scriptsize}
\tablewidth{400pt}
\tablecaption{H$_2$ properties of clumps in Orion A South}
\tablehead{
\colhead{Name\tablenotemark{a}}&
\colhead{ID}&
\colhead{H$_2$ Emission\tablenotemark{b}}\\
}

\startdata
053474-05414&01&OMC4, Jet-like, PA 40\arcdeg, Stanke 3-11\\
053514-05513&25&OMC5, Jet-like, PA 320\arcdeg, Stanke 3-16\\
053515-05558&30&OMC5, Jet-like, PA 335\arcdeg, Stanke 4-1,4-2,4-3\\
053515-05584&31&OMC5, Diffuse\\
053520-05571&37&OMC5,Jet-like, Stanke 4-4,4-5,4-6,4-7\\
053537-06131&40&OMC5, Jet-like, PA 70/250\arcdeg, Stasnke 4-11, 4-12, IRAS 05329-0614\\
053550-06269&41&HH34, Jet-like\\
053551-06265&42&HH34 IRS5, Jet-like\\
053586-06100&47&OMC5, Bubble, PA 70/250\arcdeg, Stannke 5-2, IRAS 05334-0611\\
053631-06221&52&L1641N, Jet-like, PA 355\arcdeg, HH306--310, IRAS 05338-0624\\
053631-06455&53&HH1/2, H$_2$O maser source, Jet-like, Stanke 6-6, 6-11, IRAS 05338-0647\\
053638-06461&56&HH1/2, Jet-like, Stanke 6-10, 6-15\\
053641-06226&58&L1641N, Jet-like, PA 80\arcdeg, HH 301\\
053641-06447&59&HH1/2, Jet-like, Stanke 6-12, IRAS 05339-0646\\
053660-06149&65&OMC5, Jet-like, PA 20/200\arcdeg, Stanke 5-5\\
053660-06389&66&South L1641N, Jet-like, PA 170/350\arcdeg, Stanke 5-28\\
053700-06372&70&South L1641N, Jet-like, Stanke 5-27\\
\enddata
\tablenotetext{a}{Name formed from J2000 positions (hhmm.mmdddmm.m). }
\tablenotetext{b}{Region, H$_2$ appearance, Stanke (2000) shock label, possible IRAS source}
\end{deluxetable}

\begin{table*}
\caption{Correlation between Total Flux and Presence of H$_2$ Emission}

\begin{center}
\begin{tabular}{ccccccc}
\tableline
$S_{850}$ (Jy)  & Total Clumps& Clumps with H$_2$& Percentage\cr
\tableline
\tableline
0--10    & 70        & 17& 24\cr
\cr
$>8$    & 10& 4& 40\cr 
6--8  & 10& 2& 20\cr 
4--6  & 12& 2& 17\cr
2--4  & 20& 4&20\cr
$<2$  & 19& 5&25\cr
\tableline
\end{tabular}
\end{center}
\end{table*}

\begin{table*}
\caption{Correlation between Peak Flux and Presence of H$_2$ Emission}

\begin{center}
\begin{tabular}{ccccccc}
\tableline
$S^{\rm peak}_{850}$ (Jy/bm)  & Total Clumps& Clumps with H$_2$& Percentage\cr
\tableline
\tableline
0--5   & 70        & 17& 24\cr
\cr
$>1.5$    & 3& 3& 100\cr 
1.0--1.5  & 6& 2&  33\cr 
0.5--1.0  &25& 6&  28\cr
$<0.5$    &37& 5&  15\cr
\tableline
\end{tabular}
\end{center}
\end{table*}

\begin{table*}
\caption{Correlation between Concentration and Presence of H$_2$ Emission}

\begin{center}
\begin{tabular}{ccccccc}
\tableline
$C$  & Total Clumps& Clumps with H$_2$& Percentage\cr
\tableline
\tableline
0--1    & 70        & 17& 24\cr
\cr
$>0.7$  & 5& 5& 100\cr 
0.6--0.7&12& 5& 42\cr 
0.5--0.6&10& 3& 30\cr
0.4--0.5&23& 3& 13\cr
$<0.4$  &21& 1&  5\cr
\tableline
\end{tabular}
\end{center}
\end{table*}

\clearpage 


\centering
\includegraphics[width=0.49\textwidth,clip]{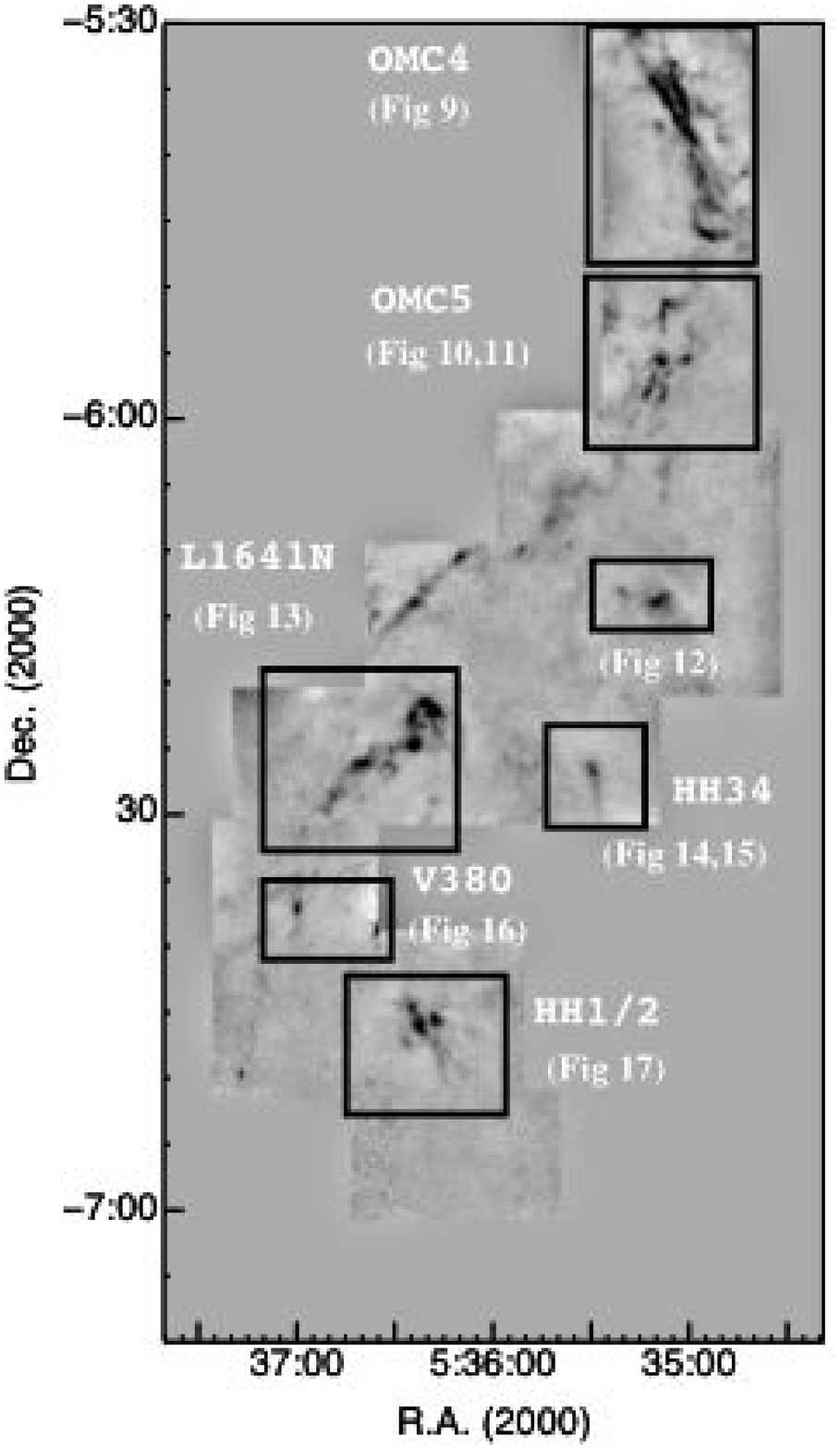}
\hfill
\includegraphics[width=0.49\textwidth,clip]{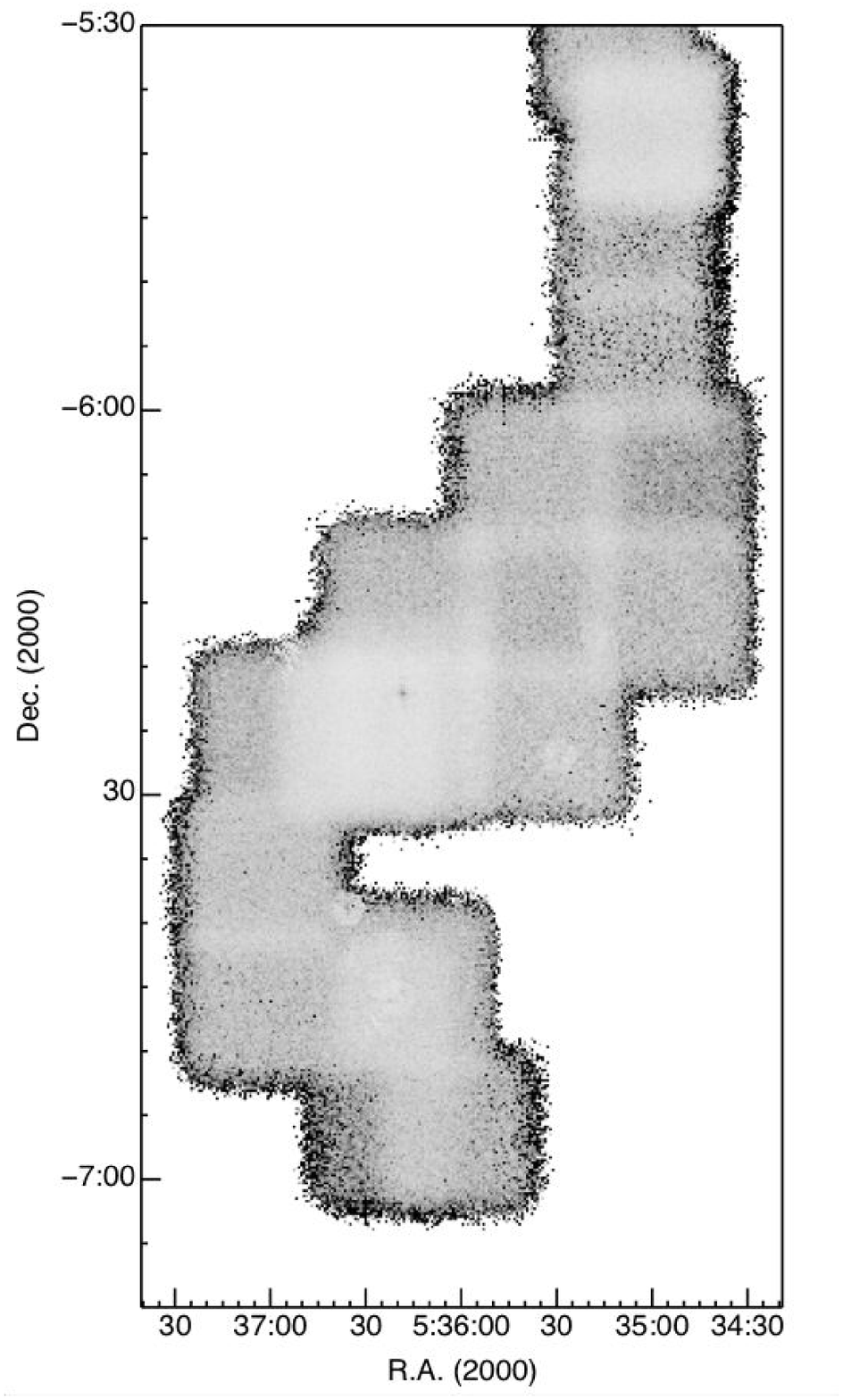}
\clearpage

\begin{figure}[htp]
\centering
\includegraphics[width=0.49\textwidth,clip]{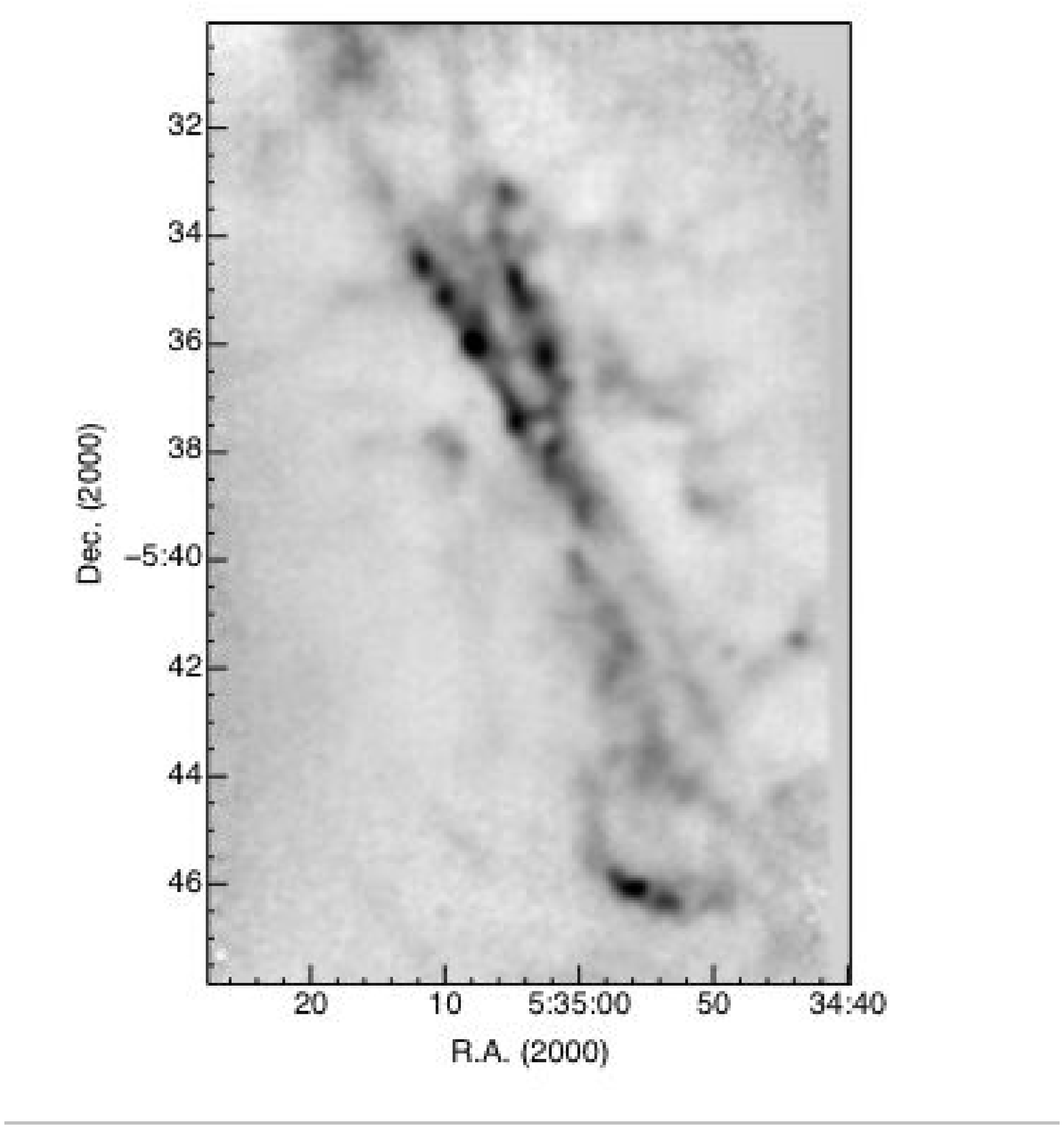}
\hfill
\includegraphics[width=0.49\textwidth,clip]{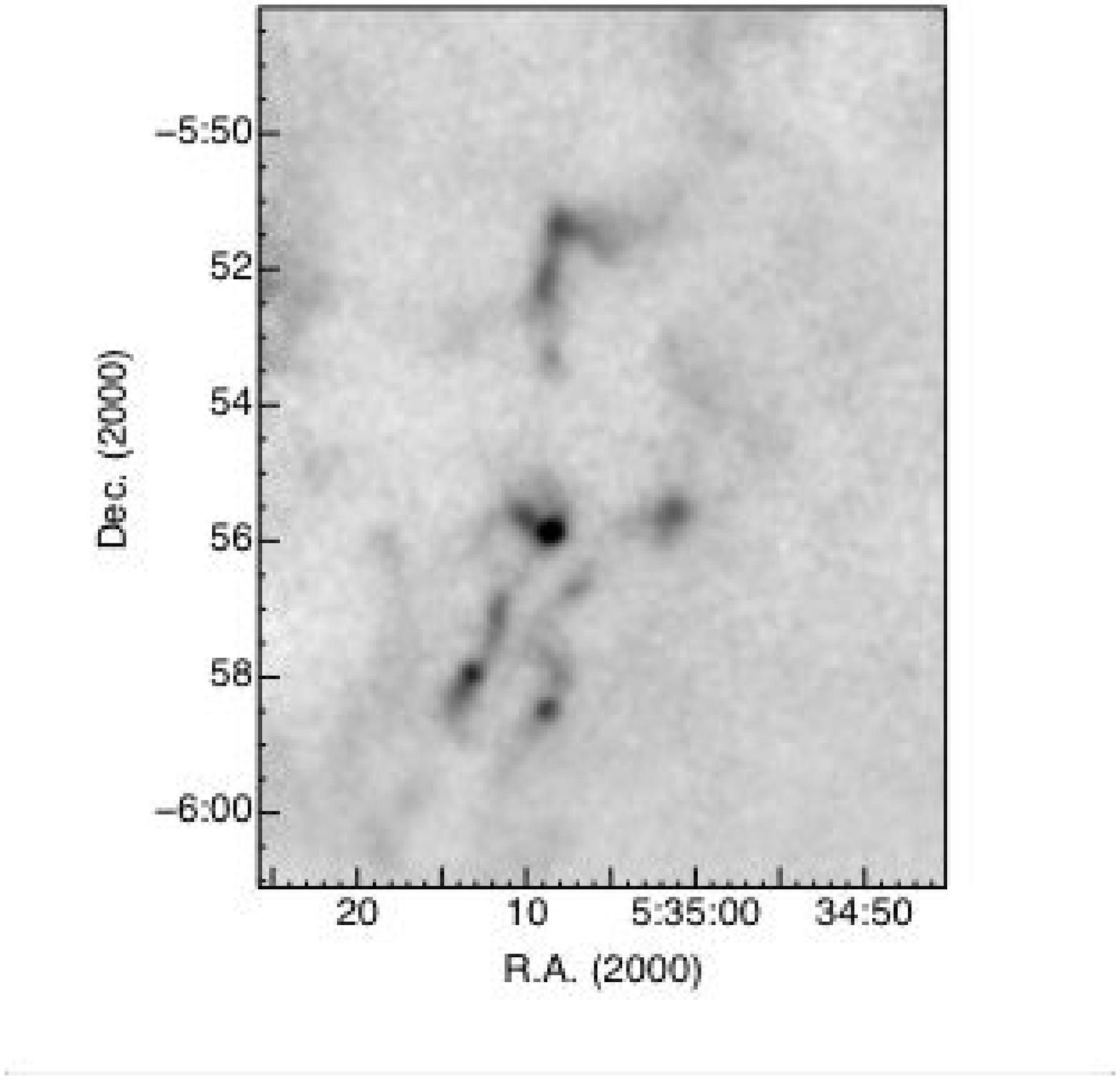}\\
\includegraphics[width=0.49\textwidth,clip]{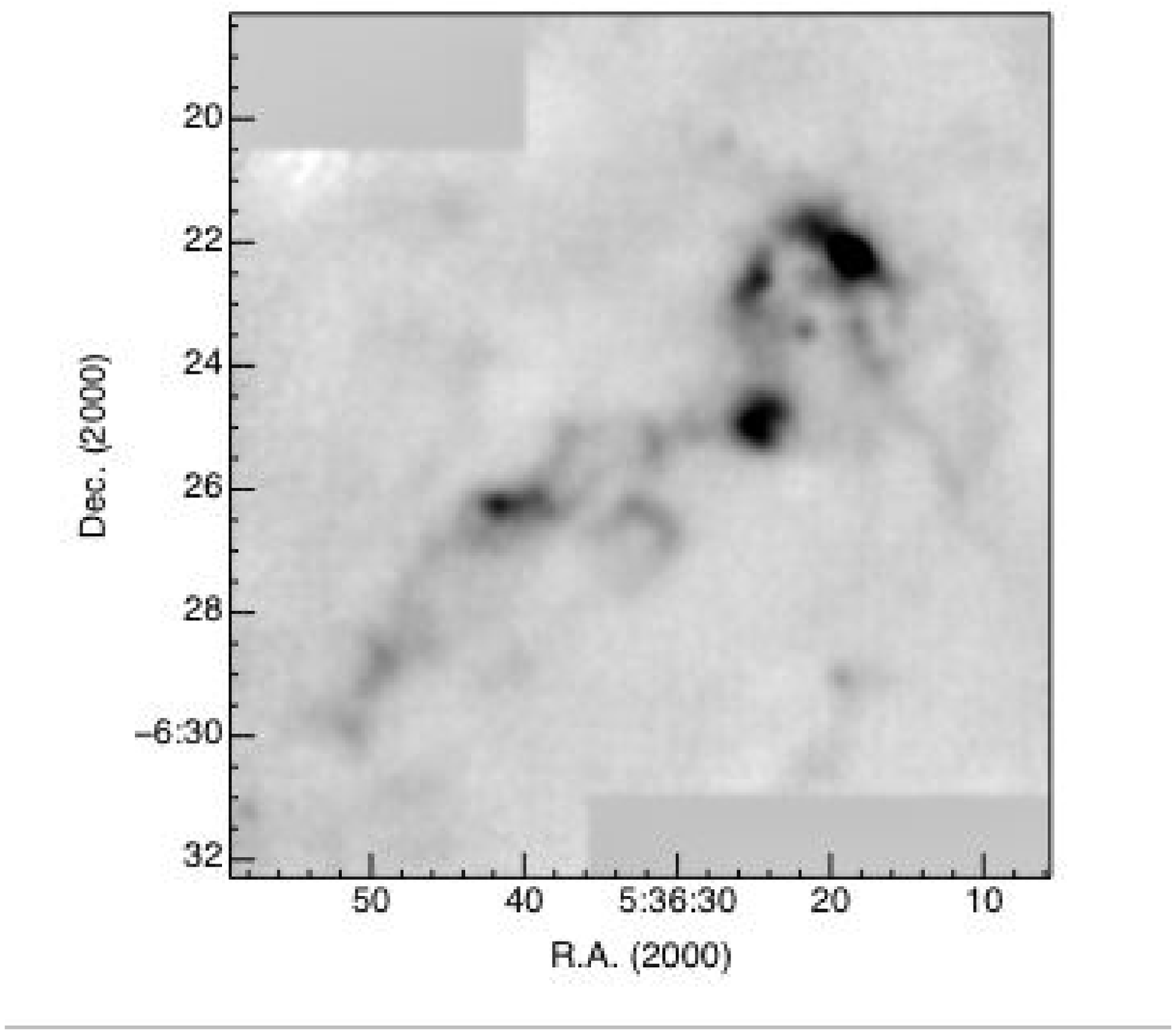}
\hfill
\includegraphics[width=0.49\textwidth,clip]{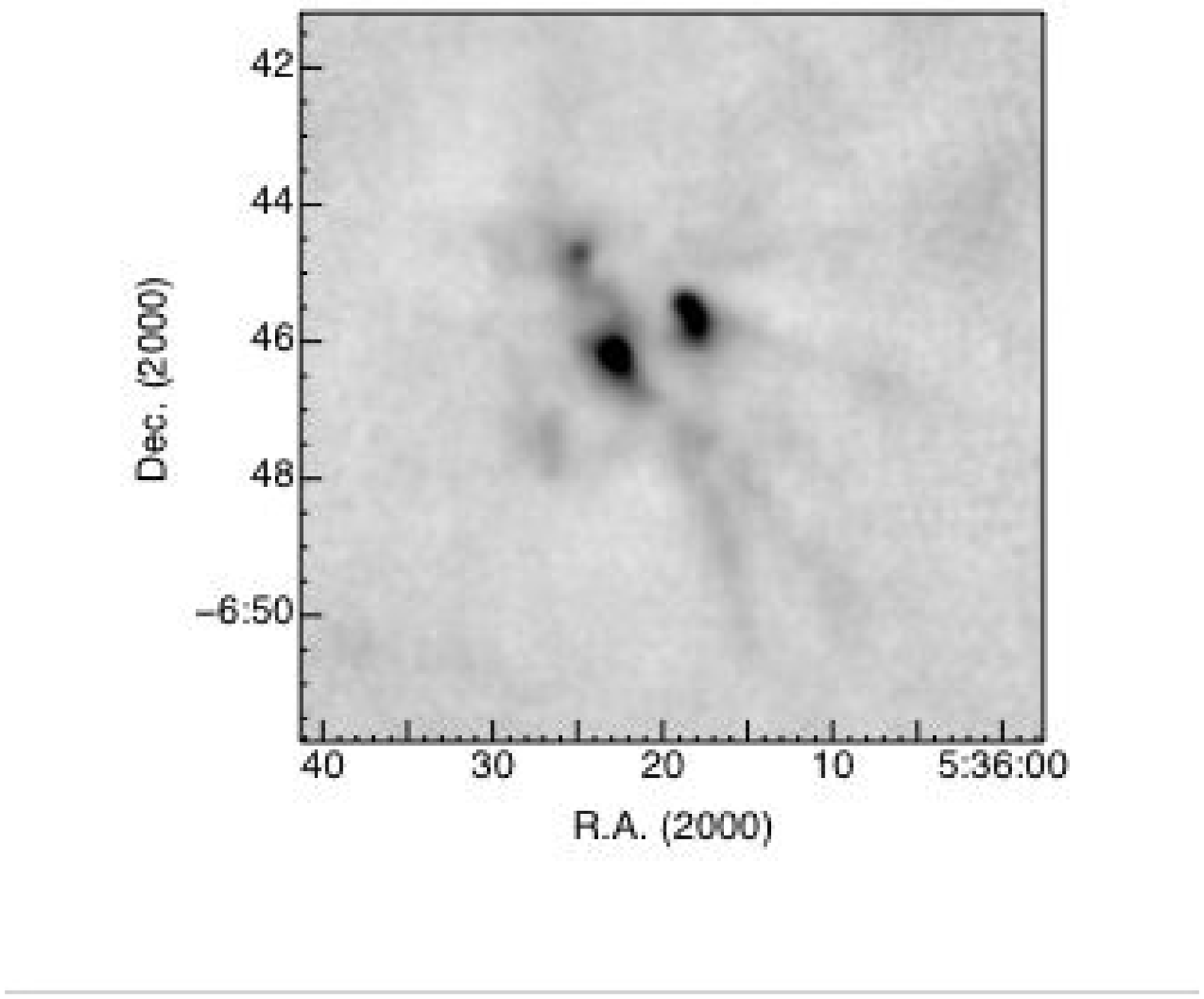}\\
\caption{Detailed maps of the regions OMC4 (Top Left), OMC5 (Top Right),
L1641N (Bottom Left), and HH1/2 (Bottom Right). In all cases the greyscale
is white to black, -0.25 to 1.0 Jy\,bm$^{-1}$.
}\label{f_oas_850b}
\end{figure}

\begin{figure}[htp]
\centering
\includegraphics[width=0.7\textwidth,clip]{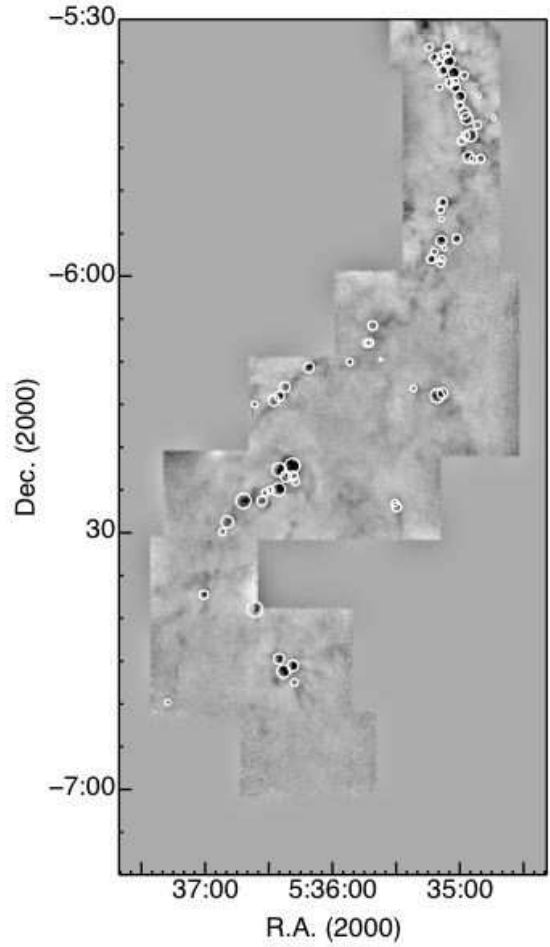}\\
\caption{Location of the submillimeter clumps found in Orion A South
850$\,\mu$m using an automated procedure (see text). The circle size
represents the area associated with each clump. 
}\label{f_oas_clumps}
\end{figure}

\begin{figure}
\includegraphics[width=0.65\textwidth,angle=90,clip]{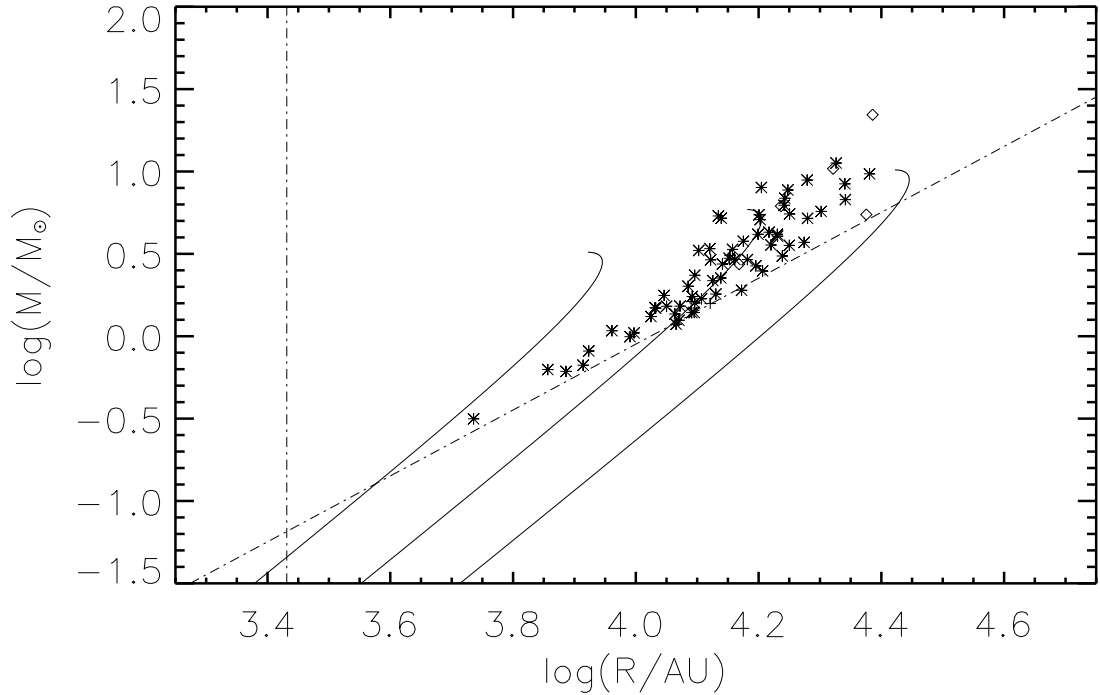}
\caption{\label{f_oas_mr}
Derived mass for each of the 71 clumps vs. the effective radius. Also plotted
are the minimum size which a clump might have (resolution limit) and the
minimum mass that a clump must have for a given radius such that the
clump finding routine can recognize it (approximately 4 $\sigma$ above the
background).  The symbols denoting each
clump are discussed in Figure \ref{f_oas_con}. The three curves
from left to right denote the mass-radius relation for Bonnor-Ebert
spheres with 20\,K internal temperatures, an additional equal internal pressure
component due to turbulence, and external pressures $P/k = 0.3, 1.0, 3.3
\times 10^6\,$K\,cm$^{-3}$ (see text).}
\end{figure}

\begin{figure}
\includegraphics[width=1.0\textwidth,angle=90,clip]{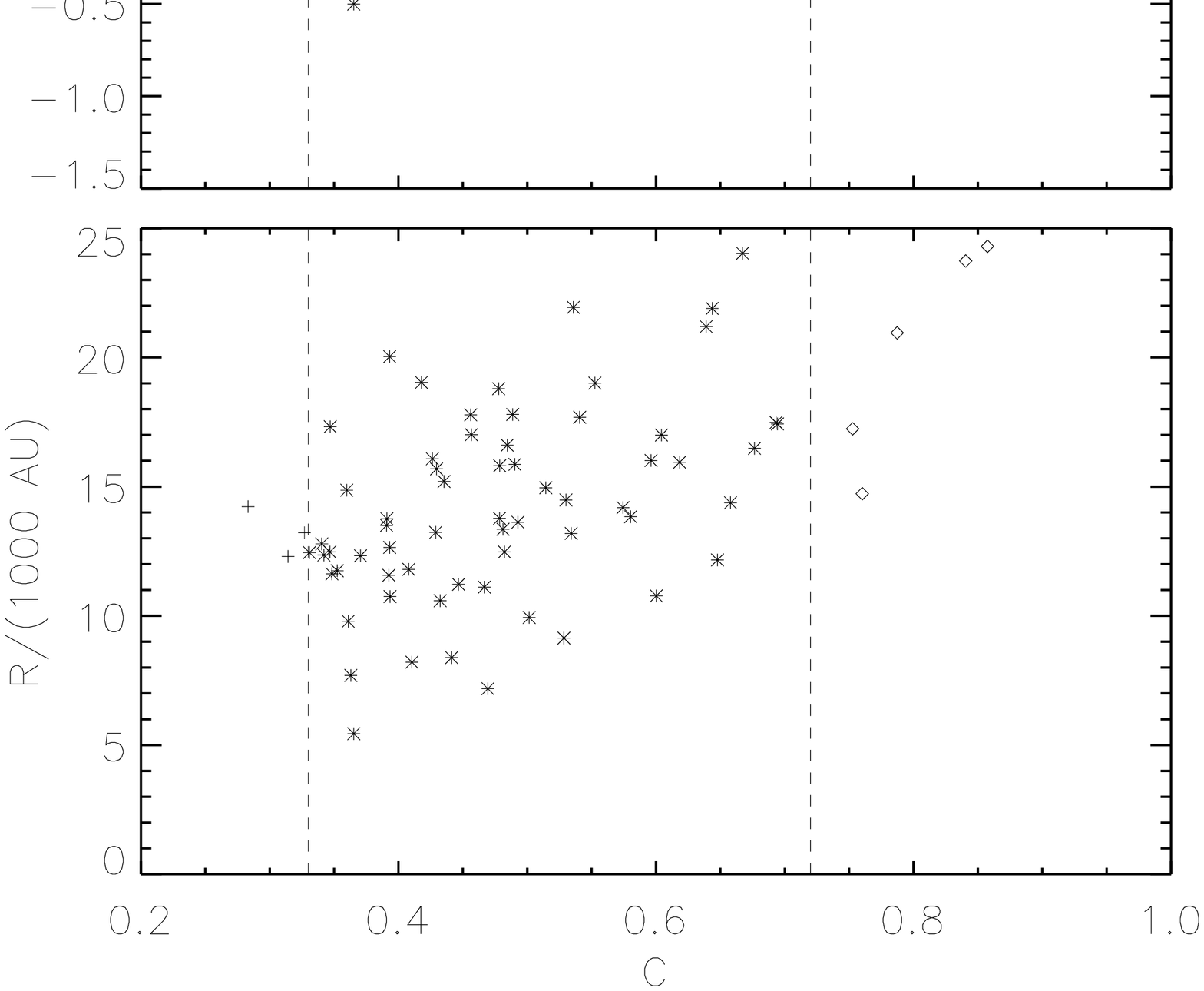}
\caption{\label{f_oas_con}
Derived concentration for each of the 71 clumps in the Orion A South
region (see text). The minimum concentration for a constant density
low-mass Bonnor-Ebert sphere is $C=0.33$, while the maximum concentration
beyond which collapse occurs is $C=0.72$. Clumps with $C > 0.72$ are denoted
by diamonds. (Top) Derived mass of the clump vs. concentration. {Note 
that the uncertainty in the mass is about 20\%, similar to the spread in
the data points.} (Bottom) Derived radius of the clump vs. concentration.}
\end{figure}

\begin{figure}
\includegraphics[width=0.65\textwidth,angle=90,clip]{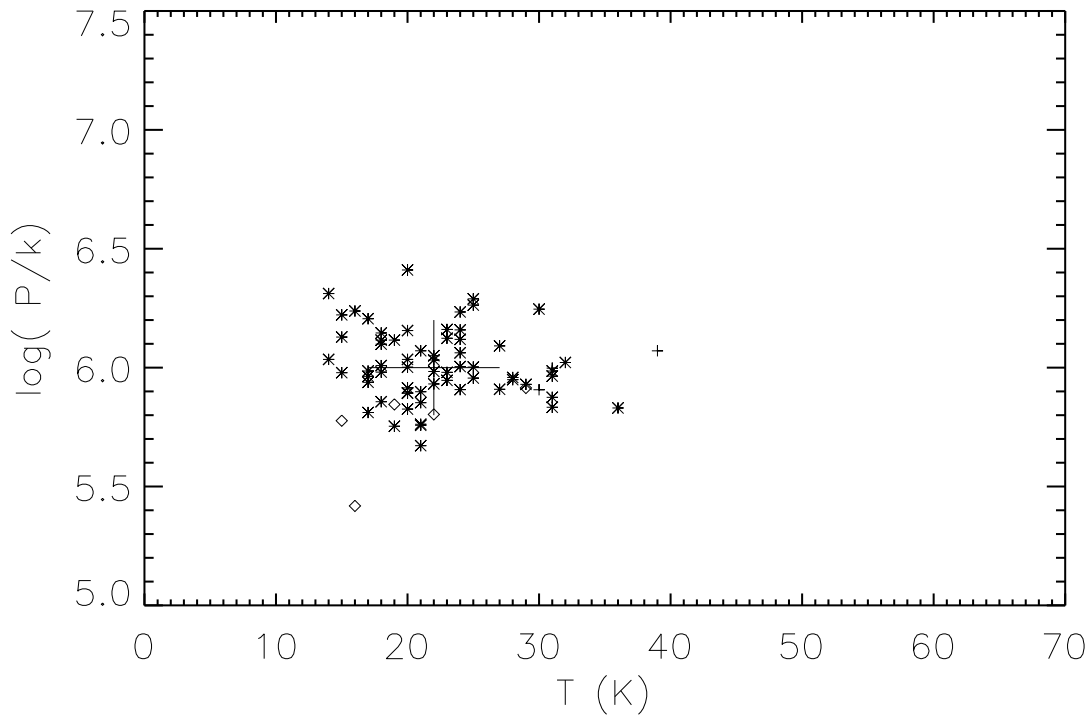}
\caption{\label{f_oas_be}
Results of determining the physical parameters of clumps using the
assumption that they are well represented by Bonnor-Ebert spheres with
measured concentrations. Plotted are the external, confining pressure vs.\
the internal temperature 
(assuming an equal contribution form turbulent support). {The cross
represents the mean and standard deviation of the data points.} The 
typical uncertainty in the derived physical parameters of an individual clump
is similar to the spread in the distribution of points.
The symbols are the same as in Fig.\ \ref{f_oas_con}.  }
\end{figure}

\begin{figure}
\includegraphics[width=0.65\textwidth,angle=90,clip]{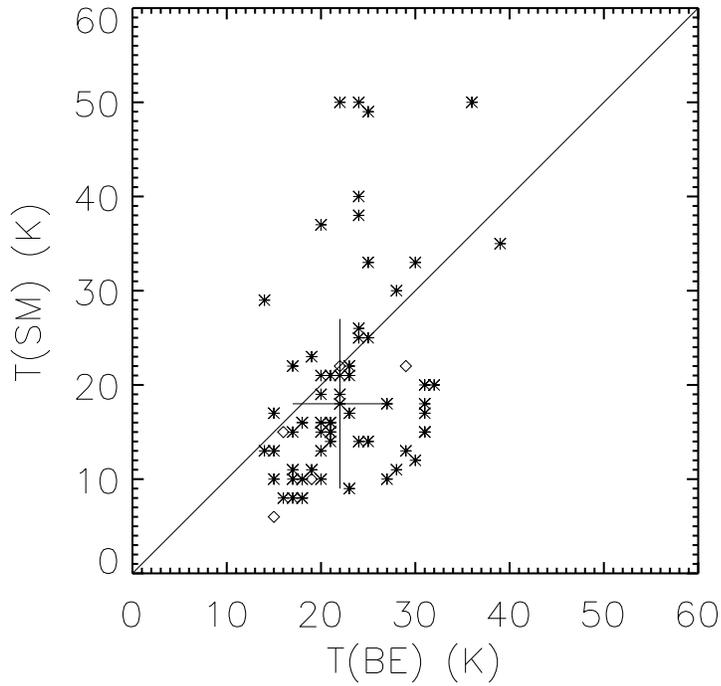}
\caption{\label{f_obs_temp}
Comparison of the clump temperature derived from the Bonnor-Ebert 
analysis versus the clump temperature derived from spectral 
energy fitting between the 450\,$\mu$m and 850\,$\mu$m observations.
A value of 50\,K is used to represent sources for which no temperature
could be determined from the spectral index.
{The cross represents the mean and standard deviation of the data points.}
}
\end{figure}

\begin{figure}
\includegraphics[width=0.65\textwidth,angle=90,clip]{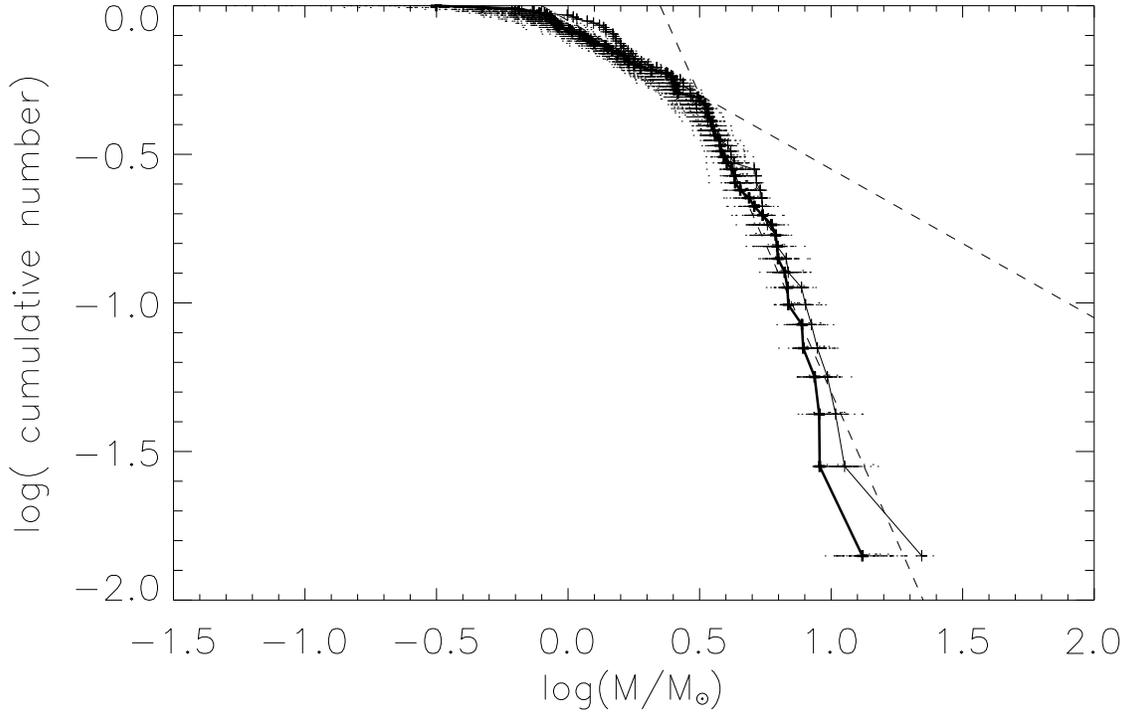}
\caption{\label{f_oas_cum}
Cumulative number function for the 850\,$\mu$m clumps in Orion A South.
The thin line converts the measured flux directly to mass using a constant
temperature $T_d = 20\,$K. The thick line converts the flux to mass using the
derived temperature from fitting the clumps to Bonnor-Ebert spheres (see text).
The horizontal lines represent the extent to which the masses might change due
to {one-sigma uncertainties} in the Bonnor-Ebert measurements. 
{Since the cumulative mass function depends on the ordering in mass of the
clumps, individual points on the horizontal lines were determined by
randomly sampling the clump distribution 100 times.} The steep 
dashed line has a slope $M^{-2}$ and approximates the high-mass end of the 
cumulative distribution.  The shallow dashed line has a slope $M^{-0.5}$ and 
approximates
the low-mass end of the distribution. The low-mass end may be severely 
incomplete (see text).}
\end{figure}

\begin{figure}
\vspace*{-20mm}
\includegraphics[width=1.\textwidth,angle=0,clip]{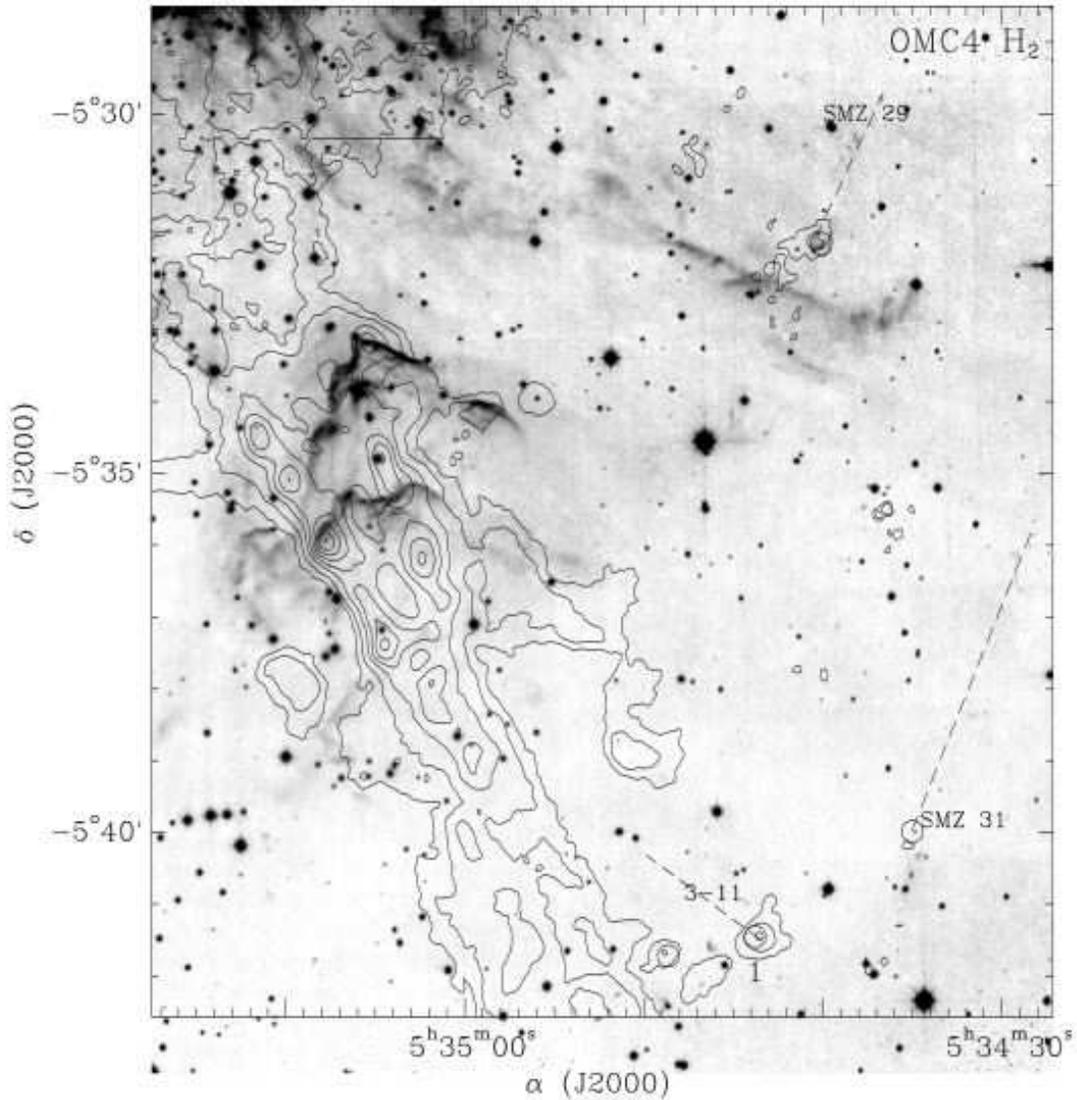}
\caption{\label{f_s30}
The OMC4 region south of the Orion Nebula.  Grey-scale shows
the H$_2$ emission from from Stanke (2000) and  Stanke, McCaughrean,
and Zinnecker (2002) overlaid on a contour map of the 850 $\mu$m
emission. In this and all subsequent figures, contour levels are 100, 
200, 400, 600, ... mJy/beam. SCUBA sources are identified by octagons and
the ID number. H$_2$ flows are identified with dashed lines and the SMZ
ID. 
}
\end{figure}

\begin{figure}
\includegraphics[width=1.\textwidth,angle=0,clip]{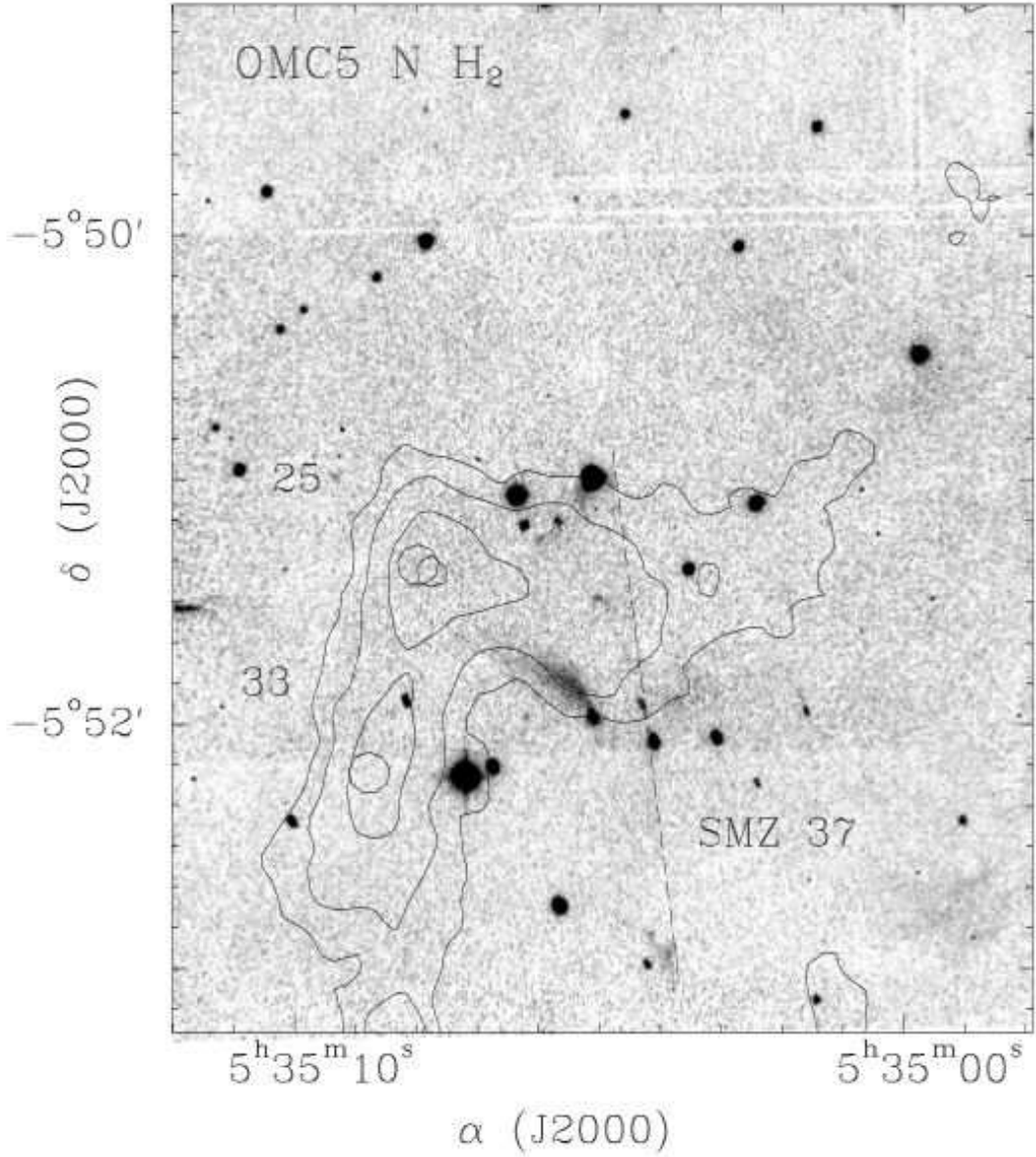}
\caption{\label{f_s40}
A small cluster of SCUBA clumps and H$_2$ jets at the northern
end of the OMC5 region located directly behind the NGC 1980
cluster in Orion. 
}
\end{figure}

\begin{figure}
\includegraphics[width=1.\textwidth,angle=0,clip]{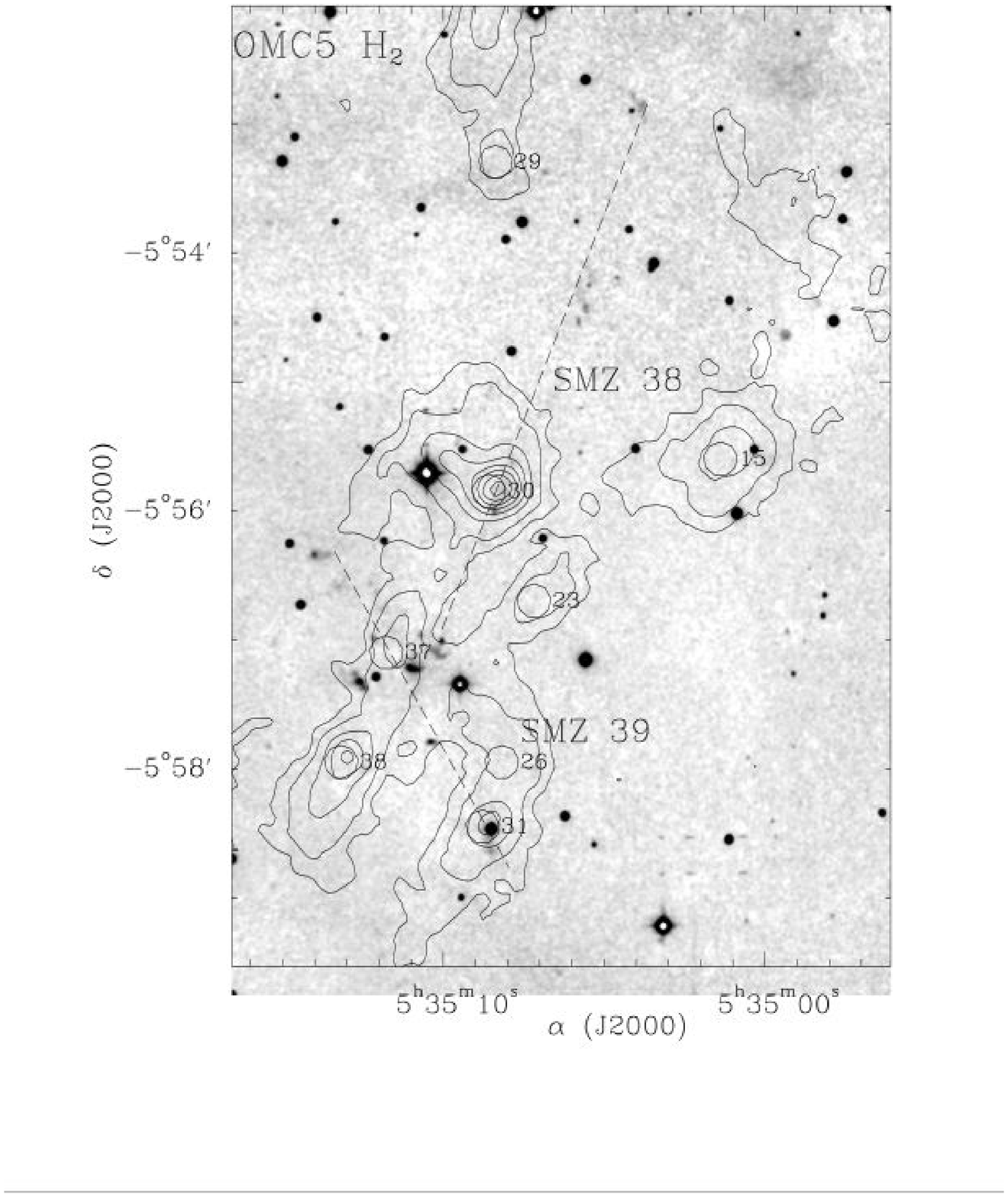}
\caption{\label{f_s1641a}
The SMZ38 and SMZ39 outflows superimposed on a contour map
of the cluster of SCUBA clumps that comprises OMC5 behind
NGC 1980.
}
\end{figure}

\begin{figure}
\includegraphics[width=.85\textwidth,angle=0,clip]{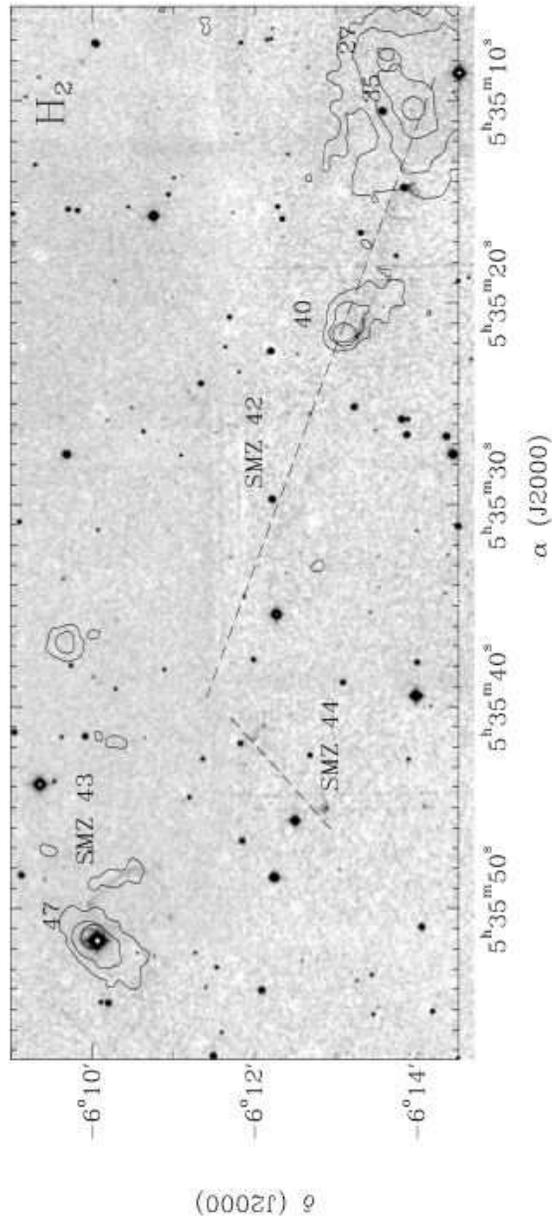}
\caption{\label{f_s1641b}
H$_2$ outflows and clumps south of NGC 1980.  An H$_2$ jet 
(SMZ42) emerges from clump ID 40 in the lower right and drives a giant
Herbig-Haro flow towards the south-west (not shown).   
A pair of H$_2$ bubbles is associated with clump ID 47 in the upper left
(SMZ43).
}
\end{figure}

\begin{figure}
\includegraphics[width=1.\textwidth,angle=0,clip]{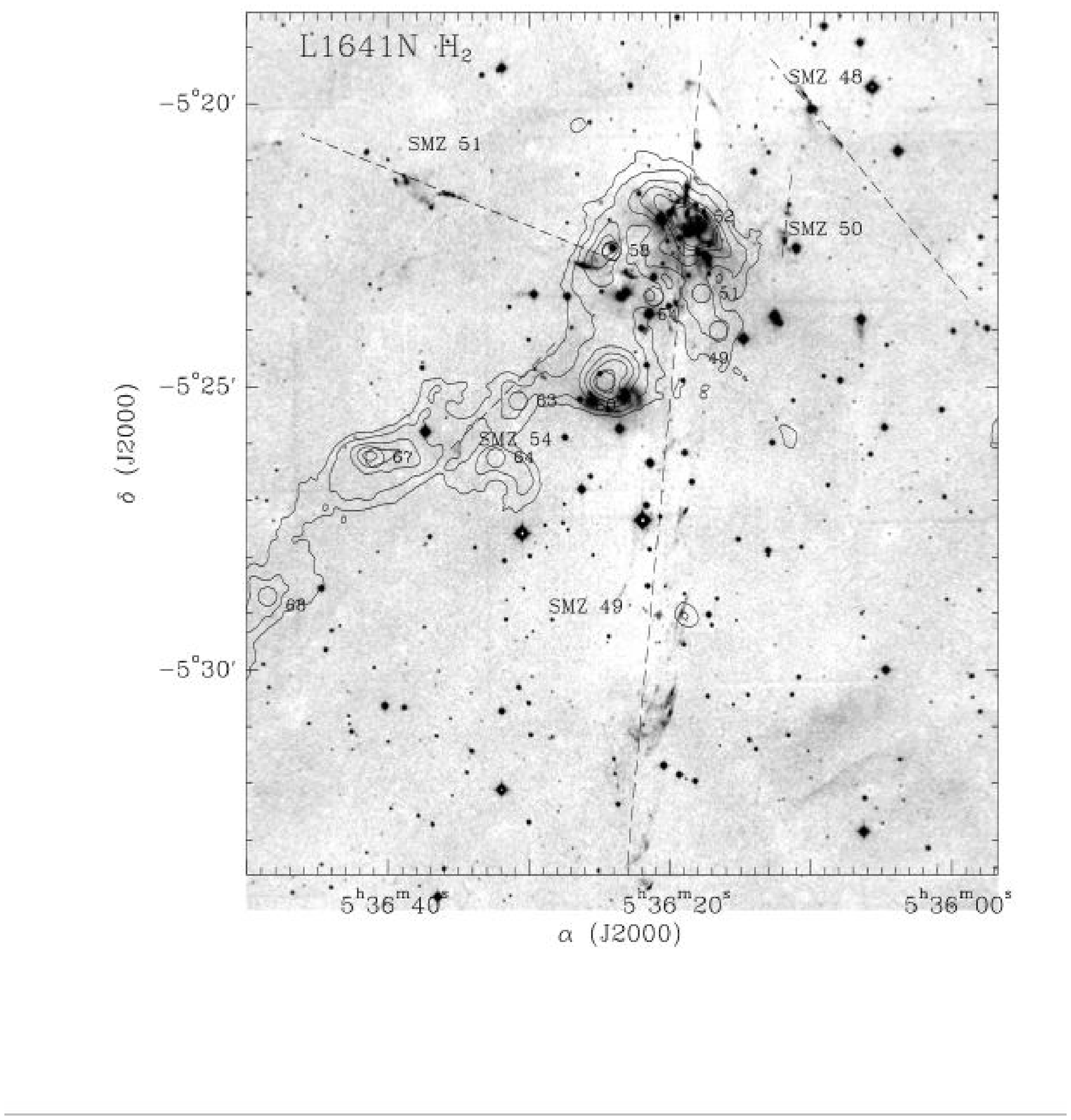}
\caption{\label{f13}
The complex L1641N region. 
}
\end{figure}

\begin{figure}
\includegraphics[width=1.\textwidth,angle=0,clip]{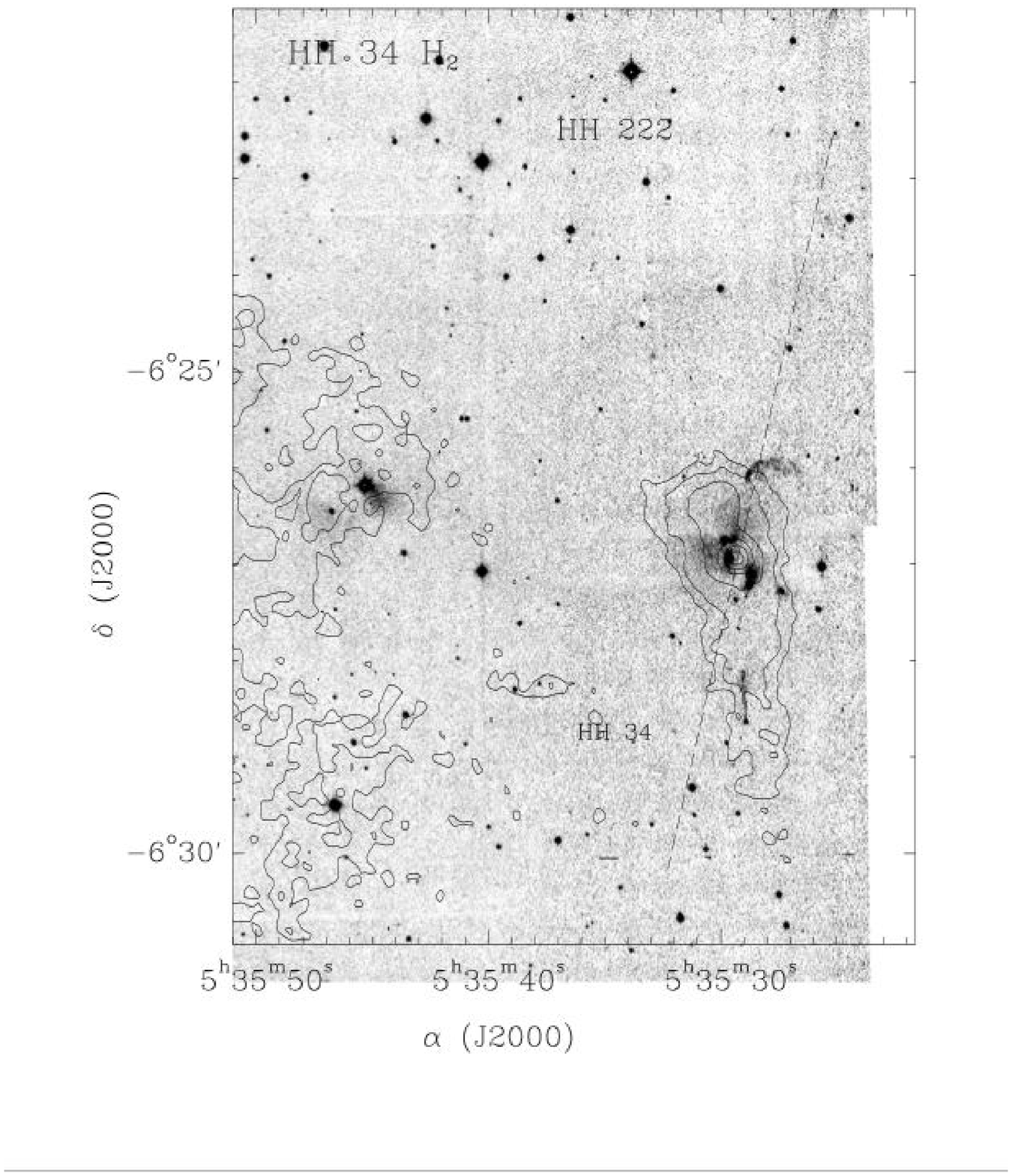}
\caption{\label{f14}
The HH 34 region showing H$_2$ emission and SCUBA contours.
SCUBA core ID 41 is the bright peak located on the axis
of the HH 34 jet (marked by the dashed line).  Core ID 42
is the protrusion that extends about 45\arcsec\ 
north-northeast of core ID 41.
}
\end{figure}

\begin{figure}
\includegraphics[width=1.\textwidth,angle=0,clip]{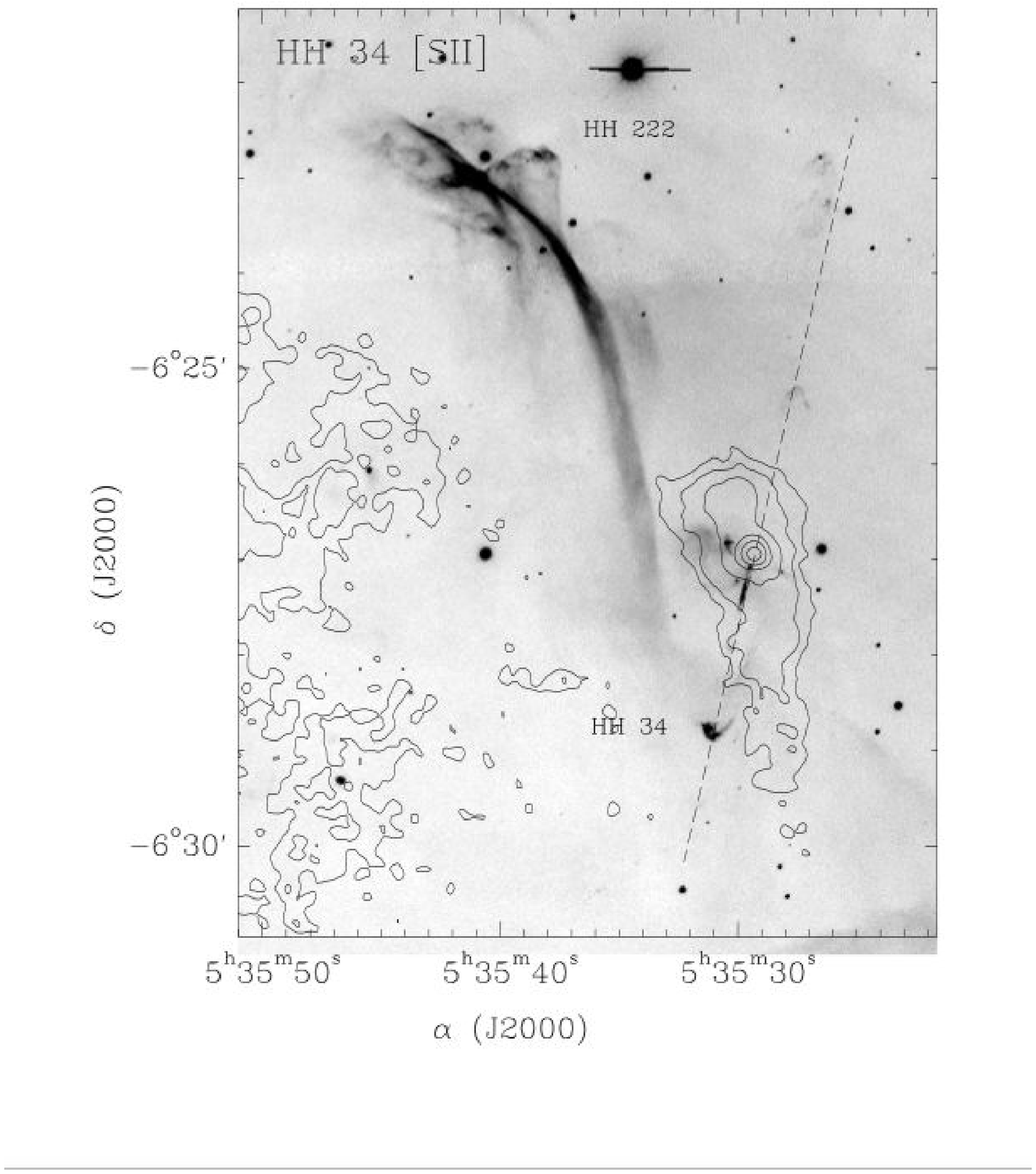}
\caption{\label{f15}
The HH 34 region showing [\sii ] emission and SCUBA contours. 
SCUBA core ID 41 is the bright peak located on the axis
of the HH 34 jet (marked by the dashed line).  Core ID 42
is the protrosion that extends about 45\arcsec\ 
north-northeast of core ID 41.
}
\end{figure}

\begin{figure}
\includegraphics[width=1.\textwidth,angle=0,clip]{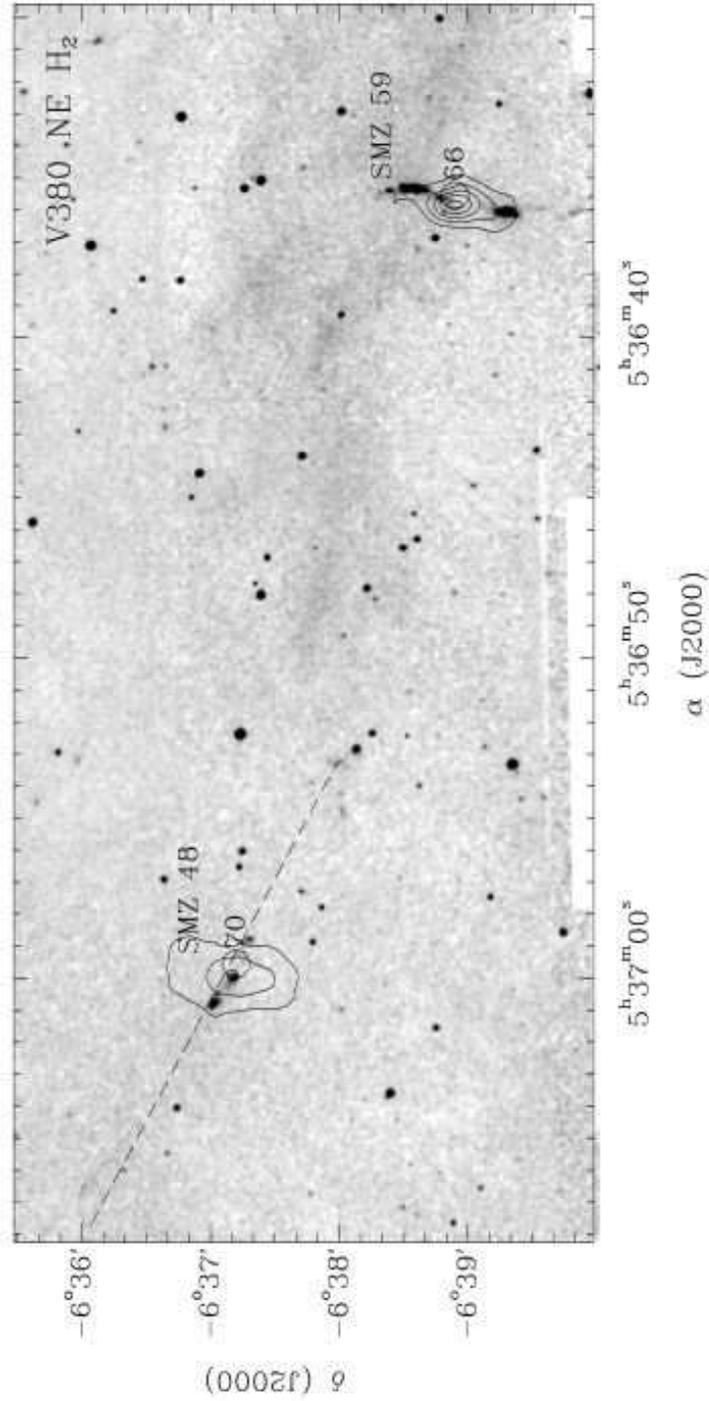}
\caption{
The S-symmetric outflow V380 Ori NE
emerging from SCUBA clump SMM 053660-06389 (ID 66, lower right).  
SCUBA clump SMM 053700-06372 (ID 70) drives a larger, 
but fainter, H$_2$ outflow (upper left).
}
\end{figure}

\begin{figure}
\vspace*{-20mm}
\includegraphics[width=1.\textwidth,angle=0,clip]{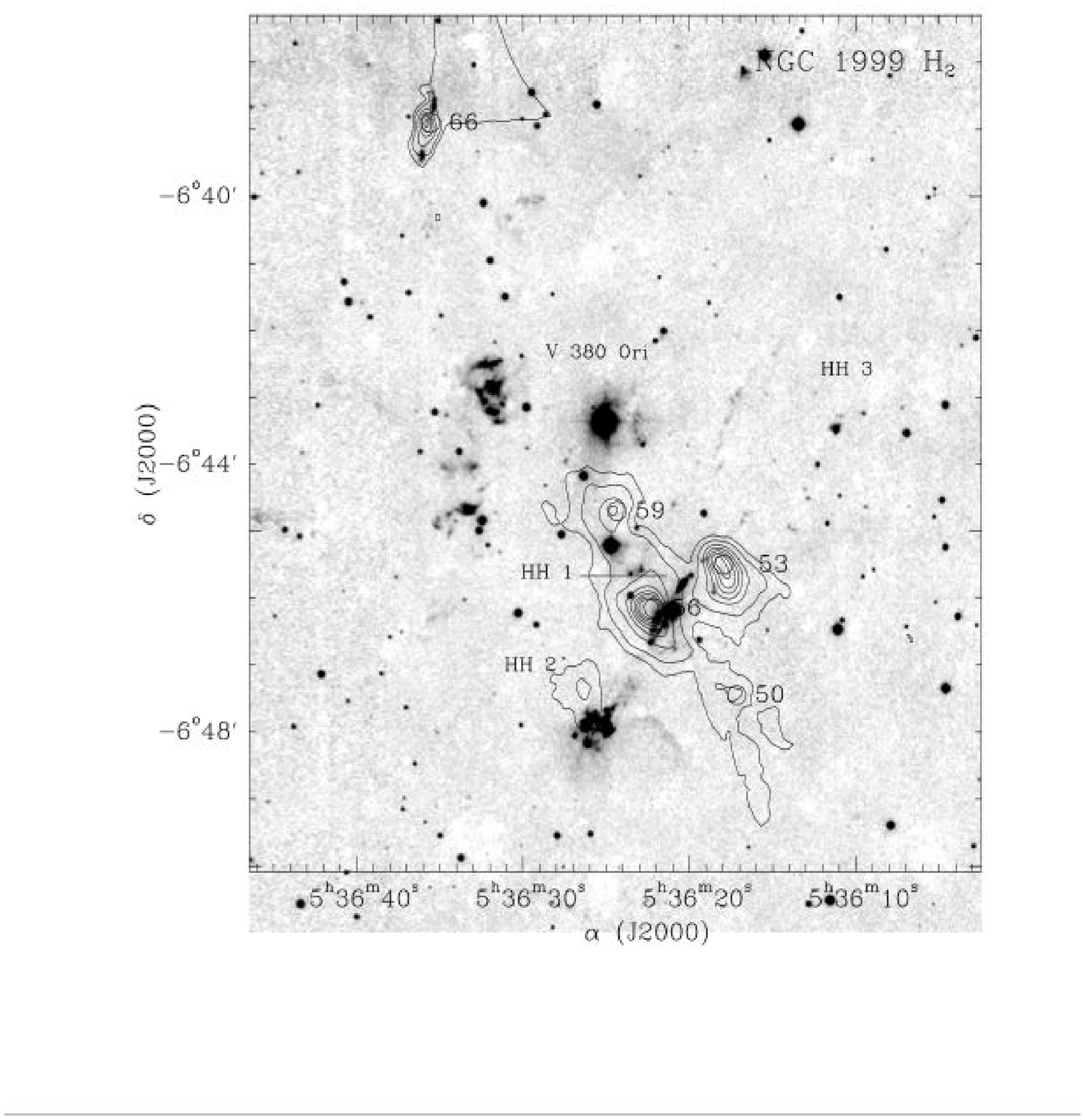}
\caption{
The NGC 1999 region containing V380 Ori (IRAS 05339-0644)
and the HH 1/2 bipolar outflow emerging from SCUBA clump SMM 053638-06461 
(ID 56).  SCUBA clump SMM 053631-06455 (ID 53) is associated with an H$_2$O 
maser and may be the driver of HH 3 in the upper right of this figure.  
SCUBA clump SMM 053641-06447 (ID 59) also drives small flow.
}
\end{figure}

\end{document}